\documentclass[
reprint,           % journal's actual layout
superscriptaddress,% authors with affiliations via superscripts
amsmath,           % add AMS-Latex features
amssymb,           % add extra AMS symbols, including amsfonts
aps,               % aps or aip
prl,               % prl, pra, prb, prc, prd, pre, prstab
notitlepage,       % control appearance of title page
longbibliography,  % show  article titles in the bibliography
floatfix,          % process floats as early as possible
nofootinbib,
]{revtex4-1}

\usepackage{tensor}     % manipulate tensors
\usepackage{graphicx}   % include figures
\usepackage[
colorlinks=true,        % color link
citecolor=blue,         % cite color
linkcolor=blue,         % link color
urlcolor=blue           % url color
]{hyperref}             % create hyperlinks
\usepackage{bm}         % \bm{<text>} Bold math symbols
\usepackage{xcolor}     % textcolor
\usepackage{lipsum}
\usepackage{color}      % \textcolor{declared-color}{text}
\usepackage[utf8]{inputenc} % accented characters for .bib file using XeLaTex
\usepackage[section]{placeins} % figures processing
\usepackage{multirow}

\newcommand{\nc}{\newcommand*}

\nc{\al}{\alpha}
\nc{\s}{\sigma}
\nc{\kp}{\kappa}
\nc{\dt}{\delta}
\nc{\Dt}{\Delta}
\nc{\Ld}{\Lambda}
\nc{\p}{\partial}
\nc{\Gm}{\Gamma}
\nc{\om}{\omega}
\nc{\Om}{\Omega}
\nc{\rd}{\mathrm{d}}
\def\({\left(}
\def\){\right)}
\def\[{\left[}
\def\]{\right]}
\def\e{\begin{equation}}
\def\q{\end{equation}}
\def\m{\begin{eqnarray}}
\def\n{\end{eqnarray}}
\nc{\Eq}[1]{Eq.~\eqref{#1}}     % equation
\nc{\Fig}[1]{Fig.~\ref{#1}}     % figure
\nc{\Table}[1]{Table~\ref{#1}}  % table
\nc{\Sec}[1]{Sec.~\ref{#1}}     % section
\nc{\Msun}{M_\odot}             % solar mass
\nc{\fpbh}{f_{\mathrm{pbh}}}    % f_pbh
\nc{\fpbhn}{f_{\mathrm{pbh0}}}    % f_pbh
\nc{\mR}{\mathcal{R}} % merger rate density
\nc{\seq}{\sigma_{\mathrm{eq}}}
\nc{\ogw}{\Omega_{\mathrm{GW}}}
\nc{\gpcyr}{\mathrm{Gpc}^{-3}\,\mathrm{yr}^{-1}}
\nc{\lvc}{LIGO/Virgo} % LIGO-VIRGO collaboration
\nc{\SNR}{\mathrm{SNR}} % signal to noise ratio
\nc{\mmin}{{m_{\mathrm{min}}}}
\nc{\mmax}{{m_{\mathrm{max}}}}
\nc{\Mmin}{{M_{\mathrm{min}}}}
\nc{\fmin}{{f_{\mathrm{min}}}}
\nc{\VT}{\mathrm{VT}}
\nc{\rhoGW}{\rho_{\mathrm{GW}}}
\nc{\vth}{\vec{\theta}}
\nc{\vd}{\vec{d}}
\nc{\vla}{\vec{\lambda}}
\nc{\Nobs}{N_{\mathrm{obs}}}
\nc{\av}[1]{\langle #1 \rangle} % average bracket
\nc{\km}{\mathrm{km}}
\nc{\Mpc}{\mathrm{Mpc}}
\nc{\Tobs}{T_{\mathrm{obs}}}
\nc{\Ntemp}{N_{\mathrm{temp}}}
\nc{\fyr}{f_{\mathrm{yr}}}
\nc{\addref}{[\textcolor{red}{add ref}] } % placeholder of references
\nc{\eg}{\textit{e.g.~}}
\nc{\app}{\approx}
\nc{\hf}{\frac{1}{2}}
\nc{\discuss}{\textcolor{red}{Add discussion here!}}
\nc{\red}[1]{\textcolor{red}{#1}}
\nc{\hp}{h_+} % h plus
\nc{\hc}{h_{\times}} % h cross
\nc{\Oh}{\hat{\Omega}}
\nc{\vx}{\vec{x}}
\nc{\mh}{\hat{m}}
\nc{\nh}{\hat{n}}
\nc{\zh}{\hat{z}}
\nc{\ph}{\hat{p}}
\nc{\A}[1]{\mathcal{A}_{#1}}
\nc{\Ogw}[1]{\Omega_{\mathrm{#1}}}
\nc{\bn}[1]{\dt\bm{t}_{\text{#1}}}
\nc{\bC}[1]{\bm{C}_{\text{#1}}}
\nc{\NTOA}{N_{\text{TOA}}}
\nc{\Nmode}{{N_{\text{mode}}}}
\nc{\ARN}{A_{\rm{RN}}}
\nc{\gRN}{\gamma_{\rm{RN}}}
\nc{\bS}{\mathbf{\Sigma}}
\nc{\br}{\mathbf{r}}
\nc{\bN}{\mathbf{R}}
\nc{\Agw}{A_\mathrm{GWB}}
\nc{\UCP}{\mathrm{UCP}}
\nc{\TT}{\mathrm{TT}}
\nc{\ST}{\mathrm{ST}}
\nc{\SL}{\mathrm{SL}}
\nc{\VL}{\mathrm{VL}}

\nc{\BFST}{$107 \pm 7$}

\begin{document}
%%%%%%%%%%%%%%%%%%%%%%%%%%%%%%%%%%%%%%%%%%%%%%%%%%%%%%%%%%%%%%%%%%%%%%%%%%%%%%%%
	
%%%%%%%%%%%%%%%%%%%%%%%%%%%%%%%%%%%% title %%%%%%%%%%%%%%%%%%%%%%%%%%%%%%%%%%%%%
\title{Tests for the existence of horizon through gravitational waves from a small binary in the vicinity of a massive object}
%\title{Gravitational waves from a compact binary revolving around an exotic compact object}
	
%%%%%%%%%%%%%%%%%%%%%%%%%%%%%%%%%%%% author %%%%%%%%%%%%%%%%%%%%%%%%%%%%%%%%%%%%
\author{Yun Fang}
\email{fang.yun@pku.edu.cn}
\affiliation{Kavli Institute for Astronomy and Astrophysics, Peking University, Beijing, 100871, China}

%%%%%%%%%%%%%%%%%%%%%%%%%%%%%%%%%%%% author %%%%%%%%%%%%%%%%%%%%%%%%%%%%%%%%%%%%
\author{Rong-Zhen Guo}
\email{guorongzhen@itp.ac.cn}
\affiliation{CAS Key Laboratory of Theoretical Physics,
	Institute of Theoretical Physics, Chinese Academy of Sciences,
	Beijing 100190, China}
\affiliation{School of Physical Sciences,
	University of Chinese Academy of Sciences,
	No. 19A Yuquan Road, Beijing 100049, China}

%%%%%%%%%%%%%%%%%%%%%%%%%%%%%%%%%%%% author %%%%%%%%%%%%%%%%%%%%%%%%%%%%%%%%%%%%
\author{Qing-Guo Huang}
\email{Corresponding author: huangqg@itp.ac.cn}
\affiliation{CAS Key Laboratory of Theoretical Physics,
Institute of Theoretical Physics, Chinese Academy of Sciences,
Beijing 100190, China}
\affiliation{School of Physical Sciences,
University of Chinese Academy of Sciences,
No. 19A Yuquan Road, Beijing 100049, China}
\affiliation{School of Fundamental Physics and Mathematical Sciences
Hangzhou Institute for Advanced Study, UCAS, Hangzhou 310024, China}
\affiliation{Center for Gravitation and Cosmology,
College of Physical Science and Technology,
Yangzhou University, Yangzhou 225009, China}
\affiliation{School of Aeronautics and Astronautics, Shanghai Jiao Tong University, Shanghai 200240, China}

%%%%%%%%%%%%%%%%%%%%%%%%%%%%%%%%%%%%% date %%%%%%%%%%%%%%%%%%%%%%%%%%%%%%%%%%%%%
\date{\today}

%%%%%%%%%%%%%%%%%%%%%%%%%%%%%%%%% abstract %%%%%%%%%%%%%%%%%%%%%%%%%%%%%%%%%%%%%
\begin{abstract}
%The nature of the horizon of exotic compact object (ECO) is unknown. Gravitational waves provide an important test for the nature of ECOs.In this work, we study the gravitational waves (GWs) of a steller-mass compact binary emitting in the vicinity of a massive ECO. The steller-mass compact binary could form in a close orbit to the supermassive compact object in galactic nuclei due to abundant astrophysical processes. Here, we assuming the centering massive object is an ECO. We obtain the GWs of the small binary by solving the Teukolsky equation numerically. We find that there exist ``continuous echo" waves propagating to infinity due to the reflecting of the GWs of the small binary by the ECO reflecting boundary, which could be comparable to the GWs of the SB emitting in a black hole background under some parameter space. These ``continuous echo" waves could provide as exquisite prob to the nature of the horizon of massive compact objects.

In this letter we calculate the gravitational waves (GWs) emitted from a small binary (SB) by solving the Teukolsky equation in the background of a massive exotic compact object (ECO) which is phenomenologically described by a Schwarzschild geometry with a reflective boundary condition at its ``would-be'' horizon. The ``continuous echo" waves propagating to infinity due to reflectivity of ECO at its ``would-be'' horizon provide an exquisite probe to the nature of the ECO's horizon.

%The existence of event horizon is the key to distinguish between a black hole and an exotic compact object (ECO). Gravitational waves (GWs) can provide valid tests for the horizon structures of ECOs. In this letter, we propose a new test to prob the event horizon of massive objects, that is through the GWs emitted by a small binary (SB) in the background of a massive ECO. Here, the massive ECO is phenomenologically described as a Schwarzschild background with a reflective boundary condition at its ``would-be'' horizon, and the GWs of the system is obtained by solving the Teukolsky equation numerically. We find that there exists ``continuous echo" waves propagating to infinity due to the reflecting of the GWs of the small binary by the ECO reflecting boundary, which could be comparable to the GWs of the SB emitting in a black hole background under some parameter space. These ``continuous echo" waves could provide as exquisite prob to the nature of the horizons of massive compact objects.
\end{abstract}

\maketitle

%%%%%%%%%%%%%%%%%%%%%%%%%%%%%%%%%%%%%%%%%%%%%%%%%%%%%%%%%%%%%%%%%%%%%%%%%%%%%%%%
\maketitle
\textbf{\textit{Introduction.}}
Black holes are predicted by general relativity as the compact stellar remnants of gravitational collapse of massive stars. Detection of event horizon, marking the boundary within which future null infinity cannot be reached, is supposed to be the decisive evidence for black hole. Both the gravitational wave (GW) detections \cite{LIGOScientific:2016aoc,LIGOScientific:2016sjg,LIGOScientific:2016dsl,LIGOScientific:2017bnn} and the observations of supermassive black holes by the Event Horizon Telescope \cite{EventHorizonTelescope:2019dse,EventHorizonTelescope:2019uob,EventHorizonTelescope:2019jan,EventHorizonTelescope:2019ths,EventHorizonTelescope:2019pgp} provide the elegant avenues toward the physics near the event horizon of black hole. However, no experiment yet is able to reveal the geometry structure near event horizon although the black hole paradigm is consistent with hitherto all electromagnetic and GW observations  \cite{LIGO_GWTC1, LIGO_GWTC2,EventHorizonTelescope:2019dse,EventHorizonTelescope:2019uob}. In particular, Barausse and Cardoso et.al.\cite{Barausse:2014tra, Cardoso:2016rao} argued that GW from a binary coalescence, especially the ringdown waveform, could not be regarded as a conclusive evidence for the existence of event horizon after the merger. Furthermore, event horizon plays a central role in the black hole information paradox \cite{Unruh:2017uaw}, and the behavior of event horizon is important to the theory of quantum gravity. Thus, probing the existence of event horizon is the cornerstone of strong-gravity research and become one of the frontier researches of fundamental physics.

On the other hand, the concept of horizonless object, namely exotic compact object (ECO), was put forward as an alternative to the classical black hole paradigm. ECO is motivated either by candidate theories of quantum gravity, including the ECO models of firewall, wormhole, fuzzball, {and black holes modified by generalized uncertainty principle }\cite{Mathur:2005zp,Mathur:2008nj,Almheiri:2012rt,Giddings_EHT2018, Bianchi_fuzzballs2020,Bena_MultiRatio2020,Mukherjee_Multimoments2020,Buoninfante:2020cqz}, or by the physics beyond standard model of particle physics, including ECO models of boson star and gravastar \cite{Mazur:2001fv,Mazur_condensestars2004, Pani_gravastar2009, Carlos_bosonstar2017}. Despite different model of ECO, it may be phenomenologically described by a standard black hole with a reflective boundary condition given at its “would-be horizon”, i.e., {its physical surface close to the event horizon predicted by general relativity}, as a result of our beliefs that the difference between ECO and black hole would be only significant in the Planckian or near-Planckian scale near the event horizon \cite{Cardoso:2019rvt}.

In literature there are several methodologies to test the nature of dark compact objects based on GW astronomy: (i) detecting the small corrections to the GW phase during binary inspiral stage, which is driven by different multipolar structures, tidal deformation, tidal heating, resonance excitation, and the correction of gravitational waves from extreme-mass-ratio inspiral systems \cite{Hughes2013, Krishnendu_2017,Cardoso_tidal2017,Maselli_Planckcorrect2018, Sennett_tidal2017, Krishnendu2019, Datta2019, Datta_tidal2019,Dreyer:2003bv,Berti:2005ys,Berti:2016lat,Berti:2006qt,Cardoso_EMRI_ECO2019,Datta_tidal2019,Maggio_EMRI_ECO2021,Sago_EMRI2021}; (ii) late-time echoes in the GW waveforms afterwards the merger stage, induced by the presence of  structure close to the gravitational radius of the ECO \cite{Cardoso_echo1st2016, Cardoso_quantumEH2016, Cardoso_2017natureAstro, Abedi2017, Conklin2018, Tsang2018, Rico2019, Nielsen_2019, Uchikata2019, Tsang2020, Xinshuo_echo2021}; (iii) stochastic GW background caused by ergoregion instability of ECOs \cite{Barausse:2018vdb,Fan:2017cfw,Du:2018cmp}. See \cite {Cardoso_review2019, Maggio_review2021} for some recent review.

\begin{figure}[t] %H
\centering
\includegraphics[height=5cm]{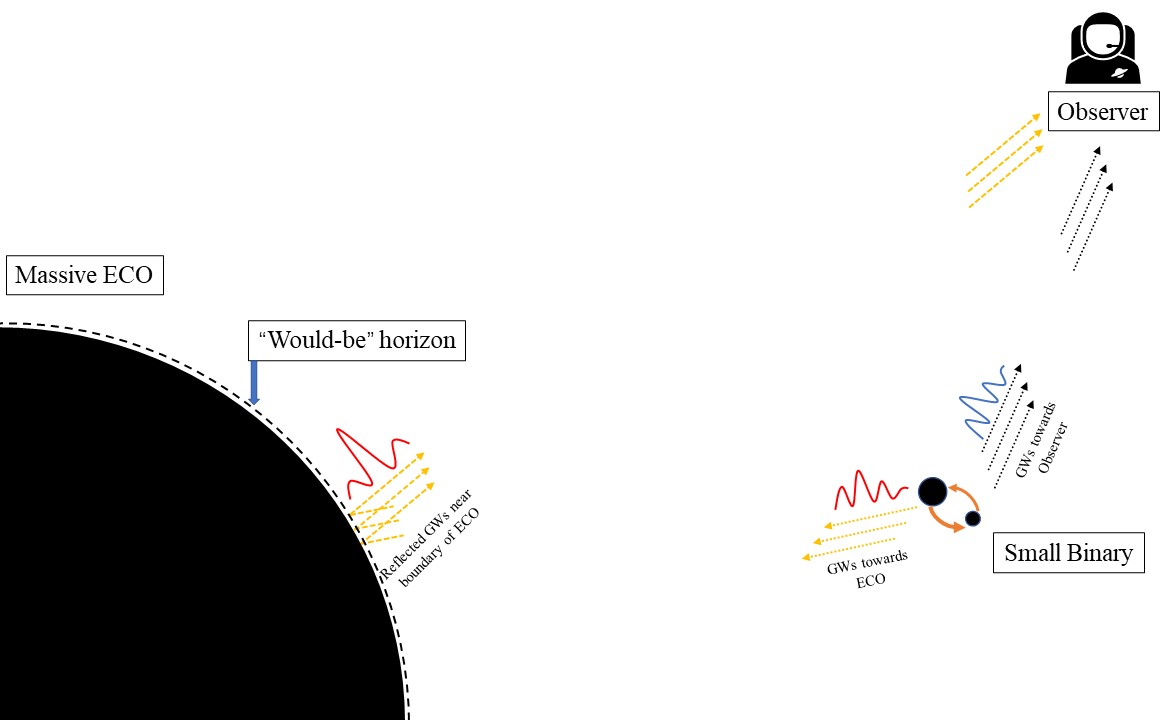}
\caption{The illustration of the GWs of a small binary emitted in the ECO background detected by a distant observer. }
\label{SB_ECO}
\end{figure}

In this letter, we propose a new method to test the existence of event horizon through GWs emitted by a small binary (SB) in the background of a massive ECO. The idea is sketched out in Fig.~\ref{SB_ECO}. Usually the Center of Galaxy harbors a supermassive black hole (SMBH) \cite{Ghez2009astro, Kormendy_2013, McConnell2013ApJ}, and SB could form in a close orbit to the SMBH due to abundant astrophysical phenomena (see, e.g., \cite{Antonini:2012ad, McKernan2012, Chen:2017xbi, Inayoshi:2017hgw, Chen:2018axp, Tagawa2020, Peng&xian2021}) and even merger in the inner-most stable circular orbit (ISCO) of the SMBH \cite{Peng&xian2021}. The orbital radius of SB is much smaller than the orbital radius of their center of mass revolving around the SMBH, and thus compose a stable hierarchical triple system, the so called ``binary-EMRI" system \cite{Chen:2018axp}.
The gravitational radiation in such system has been studied in large amount of contexts, the waveforms are characterized by many unique features \cite{Cisneros:2012sk, Meiron:2016ipr, Randall:2018lnh, Wong:2019hsq, Torres-Orjuela:2018ejx, Torres-Orjuela:2020cly, DOrazio:2019fbq,Ezquiaga:2020dao, Ezquiaga:2020gdt, Hoang:2017fvh, Randall:2019sab, Fang_2019apj, Fang_triple2020, Yuhang2020, Cardoso_tuningfork2021}.
If the centering object is a massive ECO rather a SMBH, GWs emitted by the SB will be reflected to infinity at the ``would-be" horizon of ECO, resulting the ``continuous echo" waves, and thus provides a new probe to the ``would-be" horizon of the massive ECO.

\textbf{\textit{Set up: the black hole perturbation theory.}}
In this letter, we study the gravitational radiation of an SB in the background of a massive ECO. We focus on the non-spinning ECO, and then the external space-time geometry is described by the Schwarzschild metric except imposing a reflective boundary condition at the ``would-be" horizon. The gravitational radiation of SB in such space-time can be obtained by solving the Teukolsky equation \cite{Teukolsky1973} in the standard procedure.

For the usual perturbed Schwardzschild black hole (BH), the out-going gravitational radiation is governed by
\begin{equation}
_{s}\mathcal{O}\  r^{4}\psi_{4}=4\pi r^{2}\hat{T}, \label{eq:TeukolskyPDE}
\end{equation}
where $\psi_{4}$ is the Newman-Penrose scalar describing the gravitational radiation at infinity, and the operator $_{s}\mathcal{O}$ under Schwarzschild geometry takes the form
\begin{eqnarray}\label{Teuoperator}
_{s}\mathcal{O}&=& -\left(\frac{r^{4}}{\Delta}\right)\partial_{t}^{2}
+\left(\frac{1}{\sin^{2}\theta}\right)\partial_{\varphi}^{2}
+\Delta^{-s}\partial_{r}\left(\Delta^{s+1}\partial_{r}\right) \nonumber \nonumber \\
&+&\frac{1}{\sin\theta}\partial_{\theta}\left(\sin\theta\partial_{\theta}\right)
+2s\left(\frac{i\cos\theta}{\sin^{2}\theta}\right)\partial_{\varphi} \nonumber \\
&+&2s\left(\frac{Mr^{2}}{\Delta}-r\right)\partial_{t}-s\left(s\cot^{2}\theta-1\right) \,,
\end{eqnarray}
where $M$ is the mass of SMBH, $\Delta=r^{2}-2Mr$, and $s=-2$.
Eq.~(\ref{eq:TeukolskyPDE}) can be solved by decomposing $\psi_{4}$ and the source term $\hat{T}$ into
\begin{equation}
r^{4}\psi_{4}=\int d\omega \sum_{\ell m} e^{-i\omega t}{}_{-2}Y_{\ell m}(\theta, \varphi)R_{\omega\ell m}(r)\,,
\end{equation}
\begin{eqnarray}\label{fourierT}
4\pi r^2\hat{T}=\int d\omega \sum_{\ell m} e^{-i\omega t}{}_{-2}Y_{\ell m}(\theta,\varphi)T_{\ell m\omega}(r) \,,
 \end{eqnarray}
where $\ _{-2}Y_{\ell m}$ is the spherical harmonics of spin weight $-2$. Then, Eq.~(\ref{eq:TeukolskyPDE}) is reduced to an ordinary differential equation, namely the Teukolsky radial equation, as follows:
\m \label{TRE}
%\mathbf{TRE:}
\!\left[r^2 f \frac{d^{2}}{dr^{2}}-2(r-M)\frac{d}{dr}+U(r)\right]R_{\omega\ell m}(r)=T_{\omega\ell m},
\n
where $f(r)=1-2M/r$, the potential $U(r)$ is given by
\begin{equation}
U(r)=f^{-1}\left[(\omega r)^{2}-4i\omega(r-3M)\right]-(\ell-1)(\ell+2) \,.
\end{equation}

Teukolsky radial equation can be solved by the Green function method with standard procedure. Firstly, we need two linearly independent homogeneous solutions which satisfy proper boundary conditions. For a massive BH, the two homogeneous functions are denoted by  $R_{\omega\ell m}^{\text{H}}$ and $R_{\omega\ell m}^{\infty}$ satisfying the following pure in-going condition at the event horizon and pure out-going condition at spatial infinity
\begin{equation} \label{approx_R}
\begin{array}{l}
R_{\omega\ell m}^{\text{H}}\sim \left\{ \begin{array}{ll}
B_{\omega\ell m}^{\text{hole}}\Delta^{2}e^{-i\omega r_{*}} & r\rightarrow 2M\\
r^{3}B_{\omega\ell m}^{\text{out}}e^{i\omega r_{*}}+r^{-1}B_{\omega\ell m}^{\text{in}}e^{-i\omega r_{*}} & r\rightarrow+\infty
\end{array}\right.\\
R_{\ell m\omega}^{\infty}\sim \left\{ \begin{array}{ll}
D_{\ell m\omega}^{\text{out}}e^{i\omega r_{*}}+\Delta^{2}D_{\ell m\omega}^{\text{in}}e^{-i\omega r_{*}} & r\rightarrow 2M\\
D_{\omega\ell m}^{\infty}r^{3}e^{i\omega r_{*}} & r\rightarrow+\infty
\end{array}\right.
\end{array}
\end{equation}
where $r_{*}$ is the tortoise coordinate given by $r^{*}=r+2M\ln\left|\frac{r}{2M}-1\right|$.
The Wronskian of Eq.~(\ref{TRE}) can be obtained from the asymptotic behavior at spatial infinity in Eq.~(\ref{approx_R}), namely
\m
\lim_{r \to \infty} {R^{\text{H}}_{\omega \ell m}R^{\infty \prime}_{\omega \ell m} - R^{\text{H} \prime}_{\omega \ell m}R^{\infty}_{\omega \ell m} \over r^2 f(r)}=2 i \omega B^{\text{in}}_{\omega \ell m} D^{\infty}_{\omega \ell m}\,,
\n
where the prime denotes the derivative with respect to $r$. Therefore, the Green function $G^{\text{BH}}(r,r^{\prime})$ in BH case could be constructed by
\m \label{eq:GreenBH}
&&G^{\text{BH}}\left(r,r^{\prime}\right)= \frac{1}{2i\omega B_{\omega\ell m}^{\text{in}}D^{\infty}_{\omega \ell m} } \times \\
&&\left[{\theta\left(r-r^{\prime}\right)\frac{R_{\omega\ell m}^{\text{H}}\left(r^{\prime}\right)R_{\omega\ell m}^{\infty}(r)}{r^{\prime 4}f^2(r^{\prime})}+\theta\left(r^{\prime}-r\right)\frac{R_{\omega\ell m}^{\text{H}}(r)R_{\omega\ell m}^{\infty}\left(r^{\prime}\right)}{r^{\prime 4}f^2(r^{\prime})}}\right]\,,\nonumber
\n
where $\theta\left(r-r^{\prime}\right)$ is the Heaviside step function, and the radial function $R_{\omega \ell m}$ at spatial infinity reads
\m \label{int_Rwlm}
R^{\text{BH}}_{\omega \ell m}(r)&=&\int_{2 M}^{\infty}d r^{\prime} {G^{\text{BH}}(r,r^{\prime})T_{\omega \ell m}(r^{\prime})} \nonumber \\
&=& {R^{\infty}_{\omega \ell m}(r) \over 2 i \omega B^{\text{in}}_{\omega \ell m} D^{\infty}_{\omega \ell m}}\int_{2M}^{\infty}d r^{\prime} {R^{\text{H}}_{\omega \ell m}(r^{\prime})T_{\omega \ell m}(r^{\prime})\over r^{\prime 4}f^2(r^{\prime})} \,.
\n
%Thus, to solve $R^{\text{BH}}_{\omega \ell m}$ the idea is to obtain the homogeneous solution $R^{\text{H}}_{\omega \ell m}$.

In order to get the homogenous solution %$R^{\text{H}}_{\omega \ell m}(r)$
$R_{\omega \ell m}(r)$, we usually solve the so-called Regge-Wheeler equation which has a form of wave equation as follows:
\begin{equation} \label{RWeq}
\left(\frac{d^{2}}{dr^{*2}}+\omega^{2}-V(r)\right)X_{\omega\ell m}(r)=0 \,,
\end{equation}
where
\begin{eqnarray}
V(r)=\left(1-\frac{2M}{r}\right)\left(\frac{\ell(\ell+1)}{r^{2}}-\frac{6M}{r^{3}}\right) \,,
\end{eqnarray}
and %$R^{\text{H}}_{\omega \ell m}(r)$%
$R_{\omega \ell m}(r)$ is related to $X_{\omega\ell m}(r)$ by the Chandrasekhar transformation \cite{Chandrasekhar_1992book}:
\m
\label{Chandra_T}
R_{\omega\ell m}(r)&=&r^{3}\left[\frac{2}{r}\left(1-\frac{3M}{r}\right)+2i\omega\right]\left[\frac{d}{dr_{*}}+i\omega\right]X_{\omega \ell m}(r) \nonumber \\
&+& r^{3}V(r)X_{\omega\ell m}(r).
\n
%And then the asymptotic behaviors in Eq.~(\ref{approx_R}) become
The asymptotic behaviors of the Regge-Wheeler function $X_{\omega\ell m}(r)$ is then inherited from Eq.~(\ref{approx_R}) as
%\guo{And the asymptotic behaviors of the corresponding solutions $X_{\omega\ell m}^{H}$ and $X_{\ell m\omega}^{\infty}$ are inherited from Eq.~(\ref{approx_R}) as}
\begin{equation} \label{approx_X_BH}
\begin{array}{l}
X_{\omega\ell m}^{H}\sim \left\{ \begin{array}{ll}
e^{-i\omega r_{*}} & r\rightarrow 2M\\
A_{\omega\ell m}^{\text{out}}e^{i\omega r_{*}}+A_{\omega\ell m}^{\text{in}}e^{-i\omega r_{*}} & r\rightarrow+\infty
\end{array}\right.\\
X_{\ell m\omega}^{\infty}\sim \left\{ \begin{array}{ll}
C_{\ell m\omega}^{\text{out}}e^{i\omega r_{*}}+C_{\ell m\omega}^{\text{in}}e^{-i\omega r_{*}} & r\rightarrow 2M\\
e^{i\omega r_{*}} & r\rightarrow+\infty
\end{array}\right.
\end{array}
\end{equation}
where $A^{\text{in}}_{\omega \ell m}$ is related to $B^{\text{in}}_{\omega \ell m}$ by
\m \label{fit_Bin_Ain}
B^{\text{in}}_{\omega \ell m}={\[12 i M \omega -(l-1)l(l+1)(l+2) \] \over 4 \omega^2}A^{\text{in}}_{\omega \ell m}.
\n

From now on, we switch to the gravitational radiation of an SB in the massive ECO background. The difference between ECO and BH is phenomenologically described by the reflective boundary condition at the ``would-be" horizon of ECO which can be well-defined in the Regge-Wheeler equation due to its standard form of wave equation. The boundary conditions of the homogeneous Regge-Wheeler functions for ECO take the form
\m \label{approx_X_ECO}
X_{\omega\ell m}^{\text{H ECO}} &\sim& e^{-i\omega r_{*}}+\mathcal{R}^{\text{ECO}}_{\omega \ell m}e^{i\omega r_{*}}  \ \ \ \  for\  r\rightarrow 2M \,,  \nonumber\\
X_{\ell m\omega}^{\infty\ \text{ECO}} &\sim& e^{i\omega r_{*}}  \ \ \ \   for\ r\rightarrow+\infty  \,,
\n
%\begin{equation} \label{approx_X_ECO}
%\begin{array}{l}
%X_{\omega\ell m}^{\text{H ECO}}\sim \left\{ \begin{array}{ll}
%e^{-i\omega r_{*}}+\mathcal{R}^{\text{ECO}}_{\omega \ell m}e^{i\omega r_{*}} & r\rightarrow 2M\\
%A_{\omega\ell m}^{\text{out ECO}}e^{i\omega r_{*}}+A_{\omega\ell m}^{\text{in ECO}}e^{-i\omega r_{*}} & r\rightarrow+\infty
%\end{array}\right.\\
%X_{\ell m\omega}^{\infty\text{ECO}}\sim \left\{ \begin{array}{ll}
%C_{\ell m\omega}^{\text{out ECO}}e^{i\omega r_{*}}+C_{\ell m\omega}^{\text{in ECO}}e^{-i\omega r_{*}} & r\rightarrow 2M\\
%e^{i\omega r_{*}} & r\rightarrow+\infty
%\end{array}\right.
%\end{array}
%\end{equation}
%
where $\mathcal{R}^{\text{ECO}}_{\omega \ell m}$ is the reflectivity which equals to zero for BH. Since the boundary condition for $X_{\omega\ell m}^{ \infty \ \text{ ECO}}$ is the same as that for BH, we have $X_{\omega\ell m}^{ \infty \ \text{ ECO}}=X_{\omega\ell m}^{{\infty}}$.

Similar to Eq.~(\ref{eq:GreenBH}), the Green function of Teukolsky radial equation for ECO can be constructed with the homogeneous functions $R^{\text{H ECO}}_{\omega\ell m}$ and $R^{\infty}_{\omega\ell m}$, and $R^{\text{H ECO}}_{\omega\ell m}$ is related to $X_{\omega\ell m}^{\text{H ECO}}$ by the Chandrasekhar transformations as well. Since both $R^{\text{H ECO}}_{\omega\ell m}$ and $R^{\text{H}/\infty}_{\omega\ell m}$ satisfy with the homogeneous Teukolsky equation, the homogeneous function $R^{\text{H ECO}}_{\omega\ell m}$ can be taken as a linear superposition of $R^{\text{H}}_{\omega\ell m}$ and $R^{\infty}_{\omega\ell m}$
\m
R^{\text{H ECO}}_{\omega\ell m}=R^{\text{H}}_{\omega\ell m}+\mathcal{K} R^{\infty}_{\omega\ell m} \,.
\n
Consequently, the homogeneous Regge-Wheeler function $X^{\text{H ECO}}_{\omega\ell m}=X^{\text{H}}_{\omega\ell m}+\mathcal{K} X^{\infty}_{\omega\ell m}$.
By demanding that $X^{\text{H ECO}}_{\omega\ell m}$ satisfies the boundary condition at the ECO ``would-be" horizon in Eq.~(\ref{approx_X_ECO}), the coefficient $\mathcal{K}$ is given by \cite{Mark:2017dnq}
\begin{equation}
    \mathcal{K}=\dfrac{\mathcal{R}^{\text{ECO}}_{\omega \ell m}\mathcal{T}^{\text{BH}}_{\omega\ell m}}{1-\mathcal{R}^{\text{ECO}}_{\omega \ell m}\mathcal{R}^{\text{BH}}_{\omega\ell m}} \,,
\end{equation}
where $\mathcal{R}^{\text{BH}}_{\omega\ell m}$ and $\mathcal{T}^{\text{BH}}_{\omega\ell m}$ are the reflection and transmission amplitude of the BH potential barrier, namely
\begin{equation}
    \mathcal{T}^{\text{BH}}_{\omega\ell m}=\dfrac{1}{C^{\text{out}}_{\omega\ell m}} ;\quad \mathcal{R}^{\text{BH}}_{\omega\ell m}=\dfrac{C^{\text{in}}_{\omega\ell m}}{C^{\text{out}}_{\omega\ell m}}.
\end{equation}
Thus, the Green function for ECO, $G^{\text{ECO}}(r,r^{\prime})$, reads
\begin{equation}
\label{Green_ECO}
    G^{\text{ECO}}(r,r^{\prime})=G^{\text{BH}}(r,r^{\prime})+G^{\text{DEF}}(r,r^{\prime}),
\end{equation}
where
\m \label{DEF_ECO}
G^{\text{DEF}}(r,r^{\prime})=\dfrac{\mathcal{K}}{2i\omega B^{\text{in}}_{\omega \ell m}D^{\infty}_{\omega \ell m}}\dfrac{R_{\omega\ell m}^{\mathrm{\infty}}(r)R_{\omega\ell m}^{\mathrm{\infty}}\left(r^{\prime}\right)}{r^{\prime 4}f^2(r^{\prime})}.
\n

Since the Green function in Eq.~(\ref{Green_ECO}) is separated into the BH part and the part related to the ECO reflectivity, the radial function of the GW emitted by the SB is also separated into
\m
R_{\omega\ell m}^{\text{ECO}}(r)=R_{\omega\ell m}^{\text{BH}}(r)+R_{\omega\ell m}^{\mathrm{DEF}}(r) \,,
\n
where
\m \label{int_Rwlm_ECO}
R^{\text{DEF}}_{\omega \ell m}(r)&\!=\! &\int_{2 M}^{\infty}d r^{\prime} {G^{\text{DEF}} (r,r^{\prime})T_{\omega \ell m}(r^{\prime})} \nonumber \\
%&\! = \!& {\mathcal{K} R_{\omega\ell m}^{\infty}(r) \over 2i\omega B^{\text{in}}_{\omega \ell m}D^{\infty}_{\omega \ell m} }\int_{2M}^{\infty}d r^{\prime} {R_{\omega\ell m}^{\mathrm{\infty}}\left(r^{\prime}\right)T_{\omega \ell m}(r^{\prime})\over r^{\prime 4}f^2(r^{\prime})}\nonumber \\
&= & \dfrac{\mathcal{R}^{\text{ECO}}_{\omega \ell m}\mathcal{R}^{\text{BH}}_{\omega\ell m}}{1-\mathcal{R}^{\text{ECO}}_{\omega \ell m}\mathcal{R}^{\text{BH}}_{\omega\ell m}}R^{\text{eff}}(r) \,,
\n
with $R^{\text{eff}}(r)$ given by
\m
R^{\text{eff}}(r)\!=\!\dfrac{R_{\omega\ell m}^{{\infty}}(r)/ C^{\text{in}}_{\omega \ell m} }{2i\omega B^{\text{in}}_{\omega \ell m}D^{\infty}_{\omega \ell m}}\!
\int_{2M}^{\infty}\! d r^{\prime} {R_{\omega\ell m}^{{\infty}}(r^{\prime})T_{\omega \ell m}(r^{\prime})\over r^{\prime 4}f^2(r^{\prime})} .
\n
The function $R^{\text{DEF}}_{\omega \ell m}(r)$ describes the waveform being reflected by the ``would-be" horizon of ECO.
Here, $R^{\text{DEF}}_{\omega \ell m}(r)$ can be re-written by the geometric series of $\mathcal{R}^{\text{ECO}}_{\omega \ell m}\mathcal{R}^{\text{BH}}_{\omega\ell m}$ (which is smaller than unity). In this sense, $R^{\text{DEF}}_{\omega \ell m}(r)$ can be physically explained as the linear superposition of repeated $R^{\text{eff}}(r)$ with the coefficient $(\mathcal{R}^{\text{ECO}}_{\omega \ell m}\mathcal{R}^{\text{BH}}_{\omega\ell m})^n$, which leads to the phenomenon called ``echo'', similar to that after the ringdown signal of a merger event discussed in \cite{Mark:2017dnq}. Therefore, the GWs detected by a distant observer include both the waveforms emitted by the SB in the BH background and the ``echo" waves reflected by the ``would-be" horizon of ECO. This is just what is  illustrated in Fig.~\ref{SB_ECO}.

%For a given reflectivity coefficient $\mathcal{R}^{\text{ECO}}_{\omega \ell m}$, we could numerically obtain the ECO Green function either by solving the Regge-Wheeler equation (\ref{RWeq}) for BH case and then use Eq.~(\ref{Green_ECO}), or by directly solving Regge-Wheeler equation with an ECO boundary condition in (\ref{approx_X_ECO}). We have proofed in our numerical codes that the two different methods are equivalent.

%R_{\omega\ell m}^{\mathrm{\infty}}\left(r^{\prime}\right)

%And the numerical results for $X^{\text{H} / \infty}_{\omega \ell m}(r)$ (or $R^{\text{H} / \infty}_{\omega \ell m}(r)$) are holding for arbitrary $\omega$.

%\textit{Setting sources as a binary Black hole}
\textbf{\textit{Setting sources. }}
We model the SB as two point particles of mass $m_1$ and $m_2$, whose energy momentum tensor is given by
\m \label{define_Tuv}
T^{\mu \nu}(x)=&&m_1\int_{-\infty}^{\infty}\delta^{(4)}(x-z(\tau_1)){d z^{\mu}(\tau_1)\over d \tau_1}{d z^{\nu}(\tau_1)\over d \tau_1}d\tau_1 \nonumber \\
&&+ (1\to2),
\n
where $z_i^{\mu}=(t(\tau_i),r(\tau_i),\theta(\tau_i),\varphi(\tau_i))$ (with $i=1,2$) are the world lines of the two point particles in the SB, and the normalization condition is given by $\int \delta^{(4)}(x)\sqrt{-g}d^4 x= 1$. For simplicity, we consider these two point particles have the same mass $m_1=m_2=\mu$. We take the orbital plane for the center of mass (CM) of SB as equatorial plane.
For the two point particles of SB in the $\theta, \varphi$ plane, their positions are
\m \label{def_phi_theta}
\varphi_1&=&\Omega_{\text{CM}} t + \epsilon_{\varphi} \sin(\omega_0 t),
\varphi_2=\Omega_{\text{CM}} t - \epsilon_{\varphi} \sin(\omega_0 t), \nonumber \\
\theta_1&=&\pi/2 + \epsilon_{\theta} \cos(\omega_0 t),
\theta_2=\pi/2 - \epsilon_{\theta} \cos(\omega_0 t),
\n
where $\epsilon_\varphi=\delta R_{\varphi}/ R$ and $\epsilon_\theta=\delta R_{\theta}/ R$, $R$ is the circular orbital radius of CM, $\delta R_{\varphi}$ and $\delta R_{\theta}$ are the separations of the two point particles from their CM alone $\varphi$ and $\theta$ directions defined in the Schwarzschild coordinates. Thus $\epsilon_\varphi$ and $\epsilon_\theta$ characterize the orbit of the SB. And $\Omega_{\text{CM}}$ and $\omega_0$ in Eq.~(\ref{def_phi_theta}) are the angular velocities of CM and SB in the coordinate frame. The coordinate frequency $\omega_0$ is related to the SB proper oscillation frequency $\omega^{\prime}_0$ by $\omega^{\prime}_0=U^{t}_{\text{CM}}\omega_0$, where $U^{t}_{\text{CM}}=(1-3M/R)^{-1/2}$ is time-component of the CM four velocity. In the rest frame of the SB, the proper quantity $\delta R^{\prime}_{\varphi} \propto 1/(\omega^{\prime}_0)^{2/3}$ is rescaled by $\delta R^{\prime}_{\varphi} ={\delta R_{\varphi}/(1-{2M\over r})}$. We could simply set $\epsilon_\theta\gg \epsilon_\varphi$, or $\epsilon_\varphi=0$ to resemble an elliptic orbit of the SB with a large eccentricity induced by the Kozai-Lidov mechanism \cite{Lidov1962, kozai1962}, and since $\epsilon_\theta\ll 1$, we could expand the source term by $\epsilon_\theta$ and keep to the lowest order. A similar set up could be seen in \cite{Cardoso_tuningfork2021, Cardoso_lightring2021}.

\textbf{\textit{The waveforms in different ECO models.}}
\label{p_reflectivity}
First of all, in order to sketch out the main feature of the GW waveform reflected by the ``would-be" horizon of ECO, we consider a toy model in which the reflectivity is given by
\m
\mathcal{R}_{\ell m \omega} =\epsilon\ e^{-2i  b_* \omega},
\n
where $r_*=b_*$ parameterizes the position of the physical surface, and the quantity $\epsilon \in (0, 1)$ parameterizes the amplitude of reflectivity at the ECO surface. This toy model has a constant reflectivity for different frequencies.

\begin{figure}[t] %H
\centering
\includegraphics[height=8.5cm]{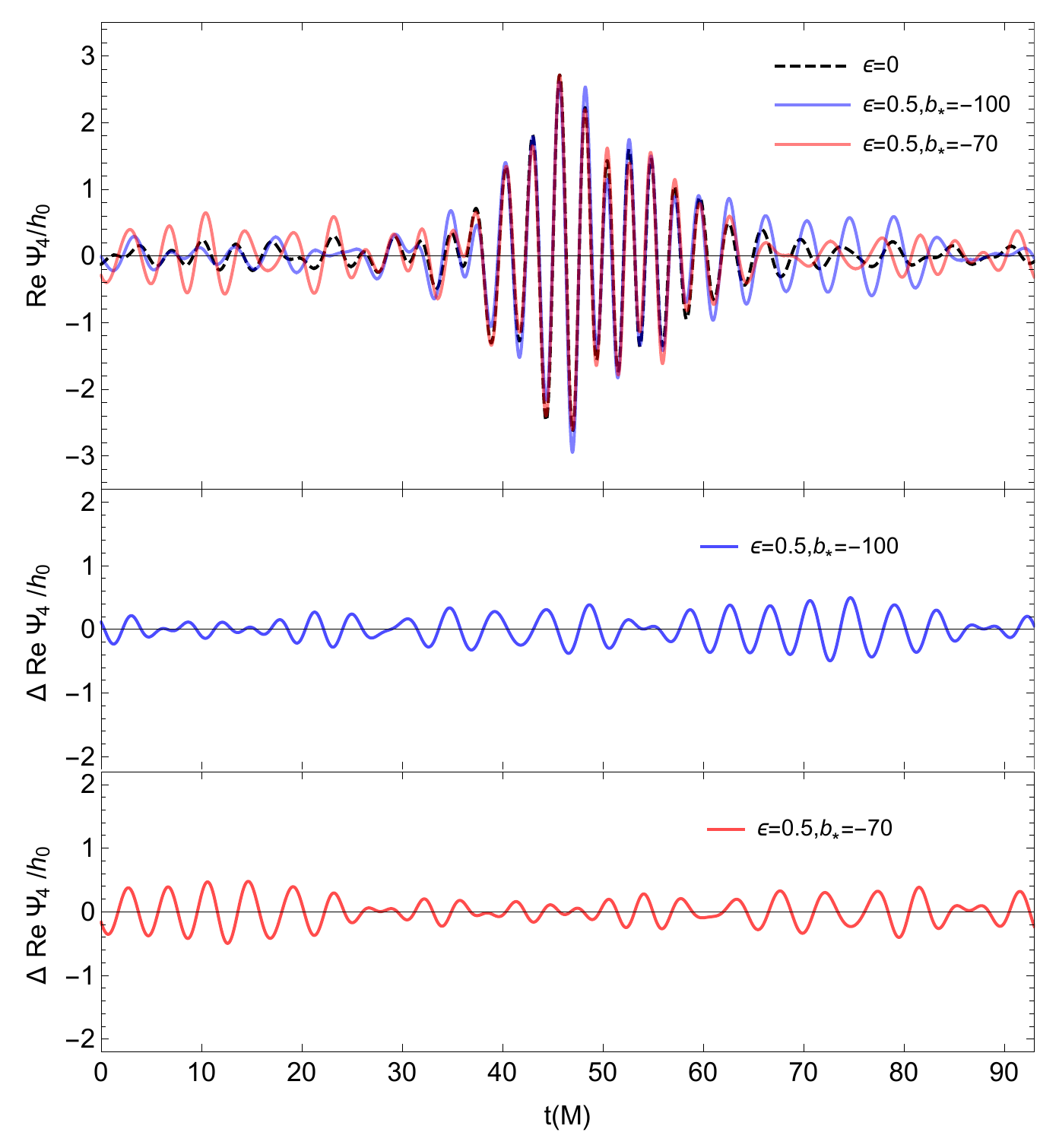}
\caption{{\bf Upper panel:} {the waveform $\psi_4$ of the SB around a massive ECO with a constant reflectivity at the surface, observed by a distant observer whose eye sight is edge on with respect to the orbital plane of the CM of the SB, i. e., the plane of $\theta=\pi/2$. The GW polarizations are related to $\psi_4$ by $\psi_{4}=\frac{1}{2}\left(\ddot{h}_{+}-i \ddot{h}_{\times}\right)$}. The CM of the SB is set in the inner-most stable circular (ISCO) of the ECO (orbital period is $93 \text{M}$), and the SB orbit has a proper frequency of {$\omega^{\prime}_0=1/\text{M}$}.
 The black dashed line with $\epsilon=0$ corresponds to the BH case. {Here, the amplitude of the waveform is rescaled by $h_0={\frac{L}{\mu ^2} \frac{\mu^{1/3}}{(2 \pi \omega^{\prime} _0)^{2/3}}}$, where $L$ is the source distance to the observer. }
{\bf Middle and Lower panels:} the difference of the SB waveforms in the ECO cases from that of the BH case, i.e. the ``echo" waveforms due to the reflection of the ECO ``would-be" horizon.
}
\label{Figure.const_reflectivity}
\end{figure}

{In Fig.~\ref{Figure.const_reflectivity}, we show the GW waveform of the SB emitted in the ECO background for an ECO under this model. The waveform is seen by a distant observer who is setting in the $\theta=\pi/2$ plane}. {From the upper panel of Fig.~\ref{Figure.const_reflectivity},  we see that the GW amplitude of the SB is magnified due to the lensing of the massive body when the SB is moving close to the back of the massive BH/ECO. Compared to the GW emitted by the SB in the BH background, the GW of the SB emitted in the ECO background is enhanced due to the reflection of GWs at the ECO ``would-be" horizon. The relative difference of the waveforms in the ECO and BH background become more obvious when the SB is not moving close to the back of the massive body, where no lensing magnification of the amplitude happens while the magnification of the GW amplitude due to the ECO reflectivity dominants}. From the Middle and Lower panels of Fig.~\ref{Figure.const_reflectivity}, we see that there are continuous GWs been reflected by the ECO surface, here, we call it ``continuous echo" waves, which has a comparable magnitude with the GWs of the SB emitting in a BH background.  {And since the two echo waves in the middle and lower panels of Fig.~\ref{Figure.const_reflectivity} are only different by $b_*$ which characteristics the echo time delay (by $2 b_*$), the only difference of the two echo waveforms is a shifting of a phase of $\Delta t=2 b_*=60 \text{M}$. }

\begin{figure}[h] %H
\centering
\includegraphics[height=6.5cm]{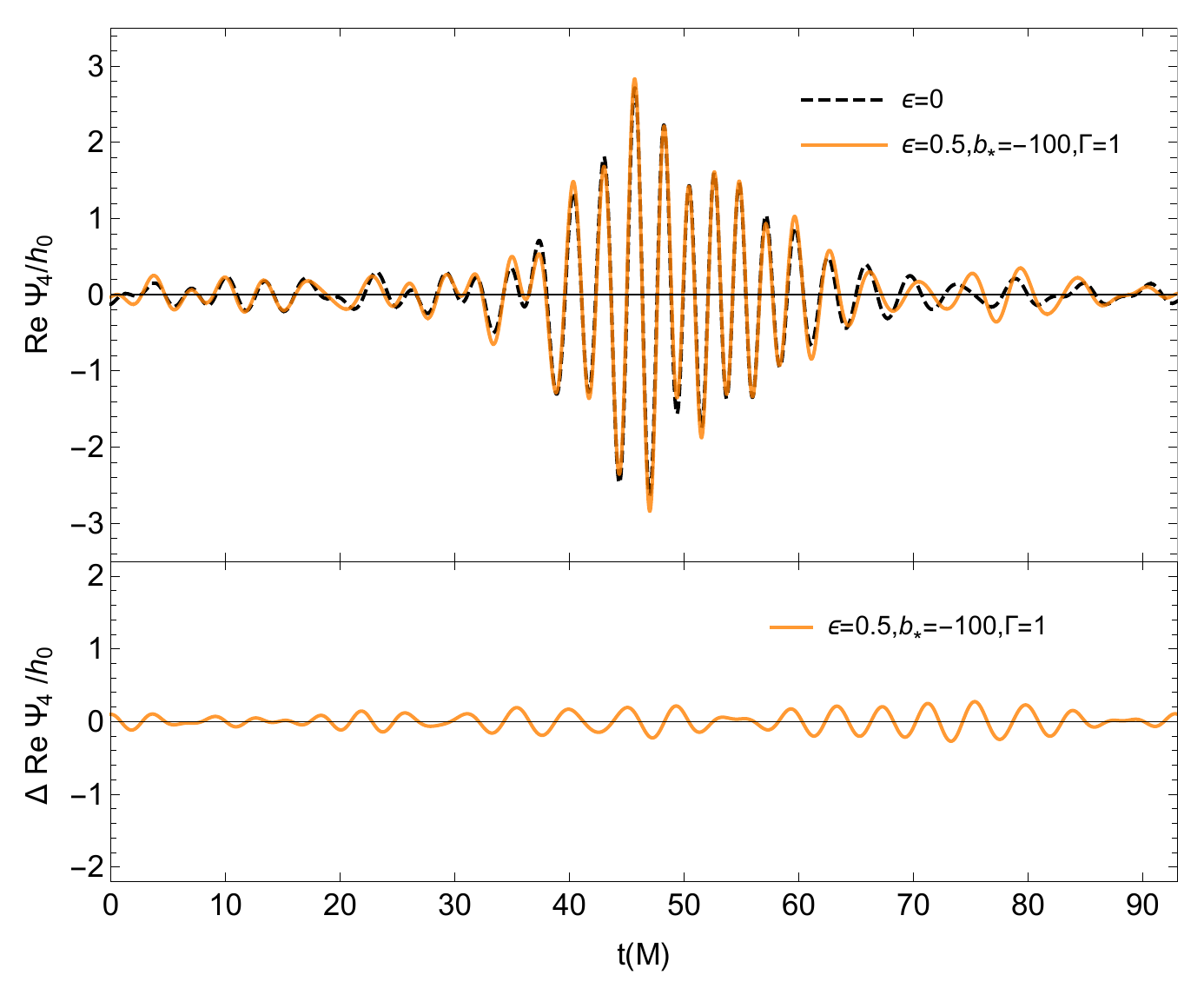}
\caption{ {\bf Upper panel:} the edge on waveform of ECO with Lorentzian reflectivity at the ECO surface, and the orbital parameters are the same as those in Fig.~\ref{Figure.const_reflectivity}.
{\bf Lower panels:} the ``echo" waveforms due to the Lorentzian reflectivity at the ECO ``would-be" horizon. }
\label{Figure.Loren_reflectivity}
\end{figure}

For a frequency-dependent reflectivity, such as Lorentzian reflectivity,  it has a {frequency dependence} as follows
\begin{equation}
     \mathcal{R}^{L}_{\ell m \omega} =\epsilon \left(\frac{ i \Gamma}{\omega +i\Gamma}\right) e^{-2i  b_* \omega}\,,
 \end{equation}
where the quantity $\Gamma$ characterizes a relaxation rate of the ECO surface and imposes a low-pass filtering of waves upon reflection in the frequency domain.  For a distant observer, GWs with frequencies  $|\omega| \lesssim \Gamma$ have the highest reflectivity. %The stability of ECO under the Lorentzian reflecitivity has been considered in \cite{Maggio:2017ivp,wang2020echoes}.
The GW waveform of the SB emitted in the ECO background with a Lorentzian reflectivity at the ECO ``would-be" horizon is shown in Fig.~\ref{Figure.Loren_reflectivity}.
 From Fig.~\ref{Figure.Loren_reflectivity}, we can see that the amplitude of the reflected waves in the Lorentzian reflectivity case is smaller than that in the constant case for the same $\epsilon$. The reason is that all the GWs of different frequencies are equally reflected by the constant reflectivity, while in the Lorentzian case only the GWs with frequency $|\omega|<{\Gamma}$ are efficiently reflected.

\begin{figure}[h] %H
\centering
\includegraphics[height=4.5cm]{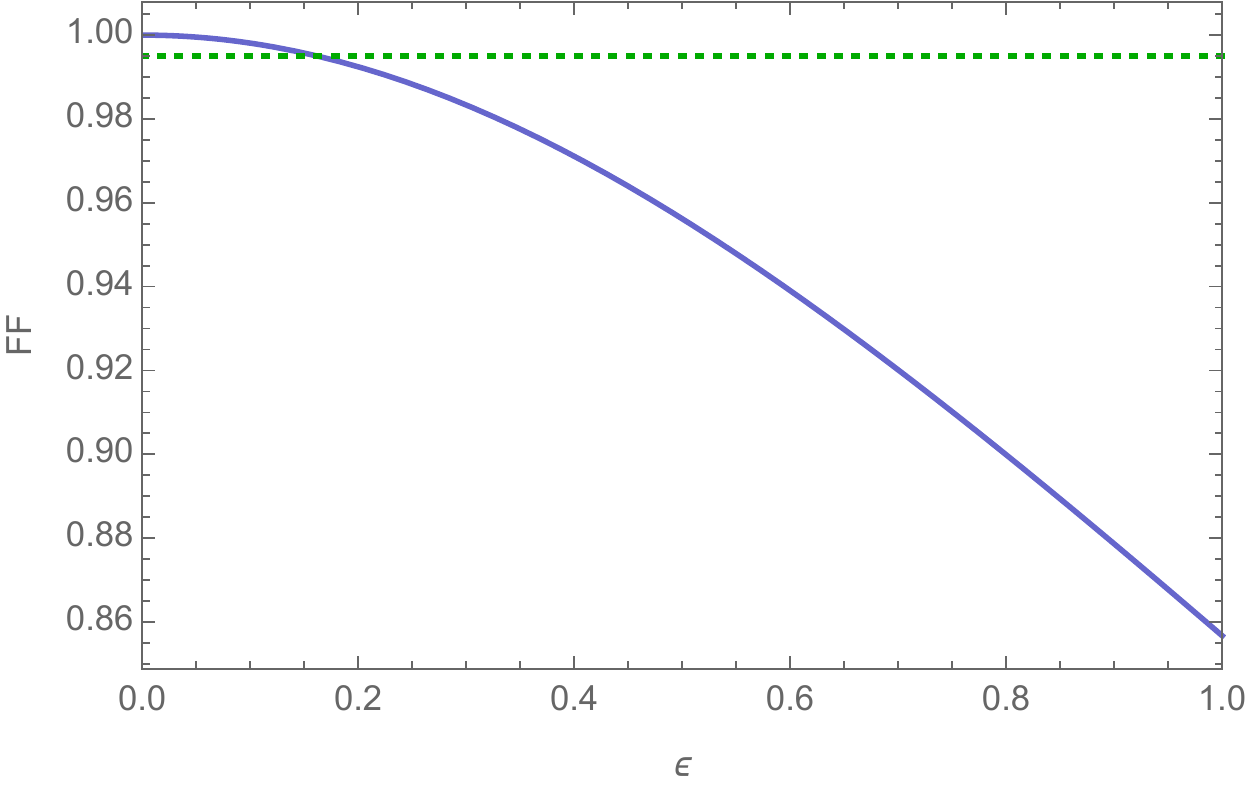}
\caption{ The fitting factor $(\textnormal{FF})$ vs. the reflective amplitude $\epsilon$ in the model with a constant reflectivity at the ECO surface. The green dashed line corresponds to $\textnormal{FF}=0.995$ which is a conservative threshold for detecting the continuous echo. 
%The fitting factor between the waveforms in the blue line (ECO case) and the black dashed line (black hole case) as shows in Fig.~(\ref{Figure.const_reflectivity}) but for a varying reflective amplitude $\epsilon$. The The green dotted line is when $\text{FFS}=0.995$, which gives the threshold for LISA's detection. 
Here, the mass of centering massive body is $10^7\Msun$ and the total mass of the binary is $40\Msun$. }
\label{FF_epsilon}
\end{figure}

\textbf{\textit{Detectability of the continuous echo.}} 
The GWs emitted by SB studied in this letter could be either LISA or LIGO-Virgo sources when they are moving during inspiral or late inspiral-merger-ringdown phase respectively, and the ``continuous echo" waves could be detectable as well. For example, we consider $\omega^{\prime}_{0}=1/M$, and the GW frequency of SB is typically in LISA band. The detectability of the ``continuous echo" waves can be revealed by comparing the GWs of the SB emitted in an ECO background with those in a black hole background. We denote the former waveform as $h_1$, and the later waveform as $h_2$. LISA employs a method called the ``matched filtering'' to discern the difference between two waveforms \citep{Finn:1992wt, Lindblom:2008cm}. The inner product of these two waveforms is defined by
\e
\langle h_1| h_2 \rangle=2 \int_{0}^{\infty} {{\tilde{h}^*_1(f)\tilde{h}_2(f)+\tilde{h}_1(f)\tilde{h}^*_2(f)} \over S_{h}(f)} df,\label{eq:inner}
\q
{where $S_{h}(f)$ is the spectral noise density \cite{Cornish:2018dyw}  and $\tilde{h}_i (f)$ denotes the Fourier transformation}
\e
\tilde{h} (f)=\int_{-\infty}^{\infty} e^{2\pi i f t}h (t)dt.
\label{fouriertrans}
\q
{These two waveforms are distinguishable if the condition}
\e
\langle \delta h|\delta h \rangle=\langle h_1-h_2| h_1-h_2 \rangle>1
\label{waveformdistincondition}
\q
is satisfied. Or equivalently, the difference between these two waveforms can be described by the
``fitting factor'' (\textnormal{FF}) \citep{Apostolatos:1994mx, Lindblom:2008cm}, namely
\e
\textnormal{FF}={\langle  h_1| h_2 \rangle \over \sqrt{ \langle  h_1| h_1 \rangle \langle  h_2| h_2 \rangle} },
\label{fittingfactor}
\q
and then one can compare it with a threshold \textnormal{FFS} defined by
\e
\textnormal{FFS}=1-{1\over \langle  h_1| h_1 \rangle+\langle  h_2| h_2 \rangle}.
\q
Here \textnormal{FFS} is closely related to the signal-to-noise ratio (${\rm SNR}=\sqrt{\langle h| h \rangle}$).  If $h_1\simeq h_2$, we
have $\textnormal{FFS} \simeq1-1/(2\,{\rm SNR}^2)$, and the criterion given in Eq.~(\ref{waveformdistincondition}) is equivalent to $\textnormal{FF}<\textnormal{FFS}$. A signal with ${\rm SNR}\geq 10$, or equivaliently $\textnormal{FFS} \geq 0.995$, is needed for claiming a detection by LISA. Therefore a conservative estimation for a distinguishable continuous echo requires $\textnormal{FF}<0.995$. The fitting factor $(\textnormal{FF})$ vs. the reflective amplitude $\epsilon$ in the model with a constant reflectivity at the ECO surface is illustrated in Fig.~\ref{FF_epsilon} which indicates that the continuous echo is detectable as long as $\epsilon>0.17$.

%Thought the detectability depends on SNR, the fitting factor do not. In Fig~(\ref{FF_epsilon}) we show the fitting factor between the waveforms in the ECO case with a constant reflectivity and the waveform in the black hole case. Form Fig~(\ref{FF_epsilon}) we see that assuming the SNR is over the threshold, the two waveforms are distinguishable for an ECO has a reflectivity coefficient of $\epsilon>0.17$ (this is the condition when FF becomes smaller than $0.995$). This fitting factor is calculated within one circle orbit motion of the center of mass of the SB around the massive body, while we did not find any significant difference assuming more orbital circles of the center of mass. Besides as LISA sources, the GWs form the SBs could also locate in LIGO/Virgo band when they are under the merger-ringdown phase,  and we leave this discussion in the future work.

%{We calculate the waveforms in Fig. \ref{Figure.const_reflectivity} and \ref{Figure.Loren_reflectivity} with Mathematica codes which we draw lessons from the published codes \cite{centra}. We have compared the our results with the results obtained from the Black Hole Perturbation Toolkit \cite{BHPtoolkit} using a semi-analytic MST method, and we find them match well. }

\textbf{\textit{Summary.}}
The SBs could form in the vicinity of a supermassive compact object in galactic nucleus due to abundant astrophysics phenomena (see, e.g., \cite{Antonini:2012ad, McKernan2012, Chen:2017xbi, Inayoshi:2017hgw, Chen:2018axp, Tagawa2020, Peng&xian2021}),  and even merger in the  ISCO orbit of the massive object \cite{Peng&xian2021}. For the SB which orbital radius is much smaller than their CM distance to the centering massive body, which satisfy $\delta R/R < (m_1+m_2)^{1/3}/M^{1/3}$, {it compose a hierarchical triple system which is stable within a relatively short time (see, e.g. \cite{Suzuki:2020vfw} for a long term stability analysis on the relativistic hierarchical triple systems)}.
In this letter, we calculate the GWs from an SB  in the vicinity of a massive ECO by solving Teukolsky equation. The reflective boundary condition at the ECO surface causes ``continuous echo" waves which last for a relatively long time until the SB merger. These ``continuous echo" waves provide an exquisite test for the nature of the ECO ``would-be" horizon.

{\it Acknowledgments. }
This work makes use of the Black Hole Perturbation Toolkit \cite{BHPtoolkit} and relativity codes provided by GRIT \cite{centra}. We also acknowledge the use of HPC Cluster of ITP-CAS and HPC Cluster of Tianhe II in National Supercomputing Center in Guangzhou. Y.F. is supported by the PKU project No. 8200904271, the fellowship of China Postdoctoral Science Foundation No. 2021M690228, and the National Science Foundation of China (NSFC) Grant No. 11721303. Q.G.H. is supported by the grants from NSFC (grant No. 11975019, 11690021, 11991052, 12047503),  the Key Research Program of the Chinese Academy of Sciences (Grant NO. XDPB15), Key Research Program of Frontier Sciences, CAS, Grant NO. ZDBS-LY-7009, CAS Project for Young Scientists in Basic Research YSBR-006, and the science research grants from the China Manned Space Project with NO. CMS-CSST-2021-B01.
	
%%%%%%%%%%%%%%%%%%%%%%%%%%%%%%%%%%%%%%%%%%%%%%%%%%%%%%%%%%%%%%%%%%%%%%%%%%%%%%%%
%%%%%%%%%%%%%%%%%%%%%%%%%%%%%%%%%%%%% references %%%%%%%%%%%%%%%%%%%%%%%%%%%%%%%%%%%%
%%%%%%%%%%%%%%%%%%%%%%%%%%%%%%%%%%%%%%%%%%%%%%%%%%%%%%%%%%%%%%%%%%%%%%%%%%%%%%%%
%\bibliographystyle{apj}
%\nocite{*}
\bibliographystyle{apsrev4-1}
\bibliography{echo}% Produces the bibliography via BibTeX.

\begin{thebibliography}{96}%
\makeatletter
\providecommand \@ifxundefined [1]{%
 \@ifx{#1\undefined}
}%
\providecommand \@ifnum [1]{%
 \ifnum #1\expandafter \@firstoftwo
 \else \expandafter \@secondoftwo
 \fi
}%
\providecommand \@ifx [1]{%
 \ifx #1\expandafter \@firstoftwo
 \else \expandafter \@secondoftwo
 \fi
}%
\providecommand \natexlab [1]{#1}%
\providecommand \enquote  [1]{``#1''}%
\providecommand \bibnamefont  [1]{#1}%
\providecommand \bibfnamefont [1]{#1}%
\providecommand \citenamefont [1]{#1}%
\providecommand \href@noop [0]{\@secondoftwo}%
\providecommand \href [0]{\begingroup \@sanitize@url \@href}%
\providecommand \@href[1]{\@@startlink{#1}\@@href}%
\providecommand \@@href[1]{\endgroup#1\@@endlink}%
\providecommand \@sanitize@url [0]{\catcode `\\12\catcode `\$12\catcode
  `\&12\catcode `\#12\catcode `\^12\catcode `\_12\catcode `\%12\relax}%
\providecommand \@@startlink[1]{}%
\providecommand \@@endlink[0]{}%
\providecommand \url  [0]{\begingroup\@sanitize@url \@url }%
\providecommand \@url [1]{\endgroup\@href {#1}{\urlprefix }}%
\providecommand \urlprefix  [0]{URL }%
\providecommand \Eprint [0]{\href }%
\providecommand \doibase [0]{http://dx.doi.org/}%
\providecommand \selectlanguage [0]{\@gobble}%
\providecommand \bibinfo  [0]{\@secondoftwo}%
\providecommand \bibfield  [0]{\@secondoftwo}%
\providecommand \translation [1]{[#1]}%
\providecommand \BibitemOpen [0]{}%
\providecommand \bibitemStop [0]{}%
\providecommand \bibitemNoStop [0]{.\EOS\space}%
\providecommand \EOS [0]{\spacefactor3000\relax}%
\providecommand \BibitemShut  [1]{\csname bibitem#1\endcsname}%
\let\auto@bib@innerbib\@empty
%</preamble>
\bibitem [{\citenamefont {Abbott}\ \emph
  {et~al.}(2016{\natexlab{a}})\citenamefont {Abbott} \emph
  {et~al.}}]{LIGOScientific:2016aoc}%
  \BibitemOpen
  \bibfield  {author} {\bibinfo {author} {\bibfnamefont {B.~P.}\ \bibnamefont
  {Abbott}} \emph {et~al.} (\bibinfo {collaboration} {LIGO Scientific,
  Virgo}),\ }\href {\doibase 10.1103/PhysRevLett.116.061102} {\bibfield
  {journal} {\bibinfo  {journal} {Phys. Rev. Lett.}\ }\textbf {\bibinfo
  {volume} {116}},\ \bibinfo {pages} {061102} (\bibinfo {year}
  {2016}{\natexlab{a}})},\ \Eprint {http://arxiv.org/abs/1602.03837}
  {arXiv:1602.03837 [gr-qc]} \BibitemShut {NoStop}%
\bibitem [{\citenamefont {Abbott}\ \emph
  {et~al.}(2016{\natexlab{b}})\citenamefont {Abbott} \emph
  {et~al.}}]{LIGOScientific:2016sjg}%
  \BibitemOpen
  \bibfield  {author} {\bibinfo {author} {\bibfnamefont {B.~P.}\ \bibnamefont
  {Abbott}} \emph {et~al.} (\bibinfo {collaboration} {LIGO Scientific,
  Virgo}),\ }\href {\doibase 10.1103/PhysRevLett.116.241103} {\bibfield
  {journal} {\bibinfo  {journal} {Phys. Rev. Lett.}\ }\textbf {\bibinfo
  {volume} {116}},\ \bibinfo {pages} {241103} (\bibinfo {year}
  {2016}{\natexlab{b}})},\ \Eprint {http://arxiv.org/abs/1606.04855}
  {arXiv:1606.04855 [gr-qc]} \BibitemShut {NoStop}%
\bibitem [{\citenamefont {Abbott}\ \emph
  {et~al.}(2016{\natexlab{c}})\citenamefont {Abbott} \emph
  {et~al.}}]{LIGOScientific:2016dsl}%
  \BibitemOpen
  \bibfield  {author} {\bibinfo {author} {\bibfnamefont {B.~P.}\ \bibnamefont
  {Abbott}} \emph {et~al.} (\bibinfo {collaboration} {LIGO Scientific,
  Virgo}),\ }\href {\doibase 10.1103/PhysRevX.6.041015} {\bibfield  {journal}
  {\bibinfo  {journal} {Phys. Rev. X}\ }\textbf {\bibinfo {volume} {6}},\
  \bibinfo {pages} {041015} (\bibinfo {year} {2016}{\natexlab{c}})},\ \bibinfo
  {note} {[Erratum: Phys.Rev.X 8, 039903 (2018)]},\ \Eprint
  {http://arxiv.org/abs/1606.04856} {arXiv:1606.04856 [gr-qc]} \BibitemShut
  {NoStop}%
\bibitem [{\citenamefont {Abbott}\ \emph {et~al.}(2017)\citenamefont {Abbott}
  \emph {et~al.}}]{LIGOScientific:2017bnn}%
  \BibitemOpen
  \bibfield  {author} {\bibinfo {author} {\bibfnamefont {B.~P.}\ \bibnamefont
  {Abbott}} \emph {et~al.} (\bibinfo {collaboration} {LIGO Scientific,
  VIRGO}),\ }\href {\doibase 10.1103/PhysRevLett.118.221101} {\bibfield
  {journal} {\bibinfo  {journal} {Phys. Rev. Lett.}\ }\textbf {\bibinfo
  {volume} {118}},\ \bibinfo {pages} {221101} (\bibinfo {year} {2017})},\
  \bibinfo {note} {[Erratum: Phys.Rev.Lett. 121, 129901 (2018)]},\ \Eprint
  {http://arxiv.org/abs/1706.01812} {arXiv:1706.01812 [gr-qc]} \BibitemShut
  {NoStop}%
\bibitem [{\citenamefont {Akiyama}\ \emph
  {et~al.}(2019{\natexlab{a}})\citenamefont {Akiyama} \emph
  {et~al.}}]{EventHorizonTelescope:2019dse}%
  \BibitemOpen
  \bibfield  {author} {\bibinfo {author} {\bibfnamefont {K.}~\bibnamefont
  {Akiyama}} \emph {et~al.} (\bibinfo {collaboration} {Event Horizon
  Telescope}),\ }\href {\doibase 10.3847/2041-8213/ab0ec7} {\bibfield
  {journal} {\bibinfo  {journal} {Astrophys. J. Lett.}\ }\textbf {\bibinfo
  {volume} {875}},\ \bibinfo {pages} {L1} (\bibinfo {year}
  {2019}{\natexlab{a}})},\ \Eprint {http://arxiv.org/abs/1906.11238}
  {arXiv:1906.11238 [astro-ph.GA]} \BibitemShut {NoStop}%
\bibitem [{\citenamefont {Akiyama}\ \emph
  {et~al.}(2019{\natexlab{b}})\citenamefont {Akiyama} \emph
  {et~al.}}]{EventHorizonTelescope:2019uob}%
  \BibitemOpen
  \bibfield  {author} {\bibinfo {author} {\bibfnamefont {K.}~\bibnamefont
  {Akiyama}} \emph {et~al.} (\bibinfo {collaboration} {Event Horizon
  Telescope}),\ }\href {\doibase 10.3847/2041-8213/ab0c96} {\bibfield
  {journal} {\bibinfo  {journal} {Astrophys. J. Lett.}\ }\textbf {\bibinfo
  {volume} {875}},\ \bibinfo {pages} {L2} (\bibinfo {year}
  {2019}{\natexlab{b}})},\ \Eprint {http://arxiv.org/abs/1906.11239}
  {arXiv:1906.11239 [astro-ph.IM]} \BibitemShut {NoStop}%
\bibitem [{\citenamefont {Akiyama}\ \emph
  {et~al.}(2019{\natexlab{c}})\citenamefont {Akiyama} \emph
  {et~al.}}]{EventHorizonTelescope:2019jan}%
  \BibitemOpen
  \bibfield  {author} {\bibinfo {author} {\bibfnamefont {K.}~\bibnamefont
  {Akiyama}} \emph {et~al.} (\bibinfo {collaboration} {Event Horizon
  Telescope}),\ }\href {\doibase 10.3847/2041-8213/ab0c57} {\bibfield
  {journal} {\bibinfo  {journal} {Astrophys. J. Lett.}\ }\textbf {\bibinfo
  {volume} {875}},\ \bibinfo {pages} {L3} (\bibinfo {year}
  {2019}{\natexlab{c}})},\ \Eprint {http://arxiv.org/abs/1906.11240}
  {arXiv:1906.11240 [astro-ph.GA]} \BibitemShut {NoStop}%
\bibitem [{\citenamefont {Akiyama}\ \emph
  {et~al.}(2019{\natexlab{d}})\citenamefont {Akiyama} \emph
  {et~al.}}]{EventHorizonTelescope:2019ths}%
  \BibitemOpen
  \bibfield  {author} {\bibinfo {author} {\bibfnamefont {K.}~\bibnamefont
  {Akiyama}} \emph {et~al.} (\bibinfo {collaboration} {Event Horizon
  Telescope}),\ }\href {\doibase 10.3847/2041-8213/ab0e85} {\bibfield
  {journal} {\bibinfo  {journal} {Astrophys. J. Lett.}\ }\textbf {\bibinfo
  {volume} {875}},\ \bibinfo {pages} {L4} (\bibinfo {year}
  {2019}{\natexlab{d}})},\ \Eprint {http://arxiv.org/abs/1906.11241}
  {arXiv:1906.11241 [astro-ph.GA]} \BibitemShut {NoStop}%
\bibitem [{\citenamefont {Akiyama}\ \emph
  {et~al.}(2019{\natexlab{e}})\citenamefont {Akiyama} \emph
  {et~al.}}]{EventHorizonTelescope:2019pgp}%
  \BibitemOpen
  \bibfield  {author} {\bibinfo {author} {\bibfnamefont {K.}~\bibnamefont
  {Akiyama}} \emph {et~al.} (\bibinfo {collaboration} {Event Horizon
  Telescope}),\ }\href {\doibase 10.3847/2041-8213/ab0f43} {\bibfield
  {journal} {\bibinfo  {journal} {Astrophys. J. Lett.}\ }\textbf {\bibinfo
  {volume} {875}},\ \bibinfo {pages} {L5} (\bibinfo {year}
  {2019}{\natexlab{e}})},\ \Eprint {http://arxiv.org/abs/1906.11242}
  {arXiv:1906.11242 [astro-ph.GA]} \BibitemShut {NoStop}%
\bibitem [{\citenamefont {Abbott}\ \emph {et~al.}(2019)\citenamefont {Abbott}
  \emph {et~al.}}]{LIGO_GWTC1}%
  \BibitemOpen
  \bibfield  {author} {\bibinfo {author} {\bibfnamefont {B.~P.}\ \bibnamefont
  {Abbott}} \emph {et~al.} (\bibinfo {collaboration} {LIGO Scientific,
  Virgo}),\ }\href {\doibase 10.1103/PhysRevX.9.031040} {\bibfield  {journal}
  {\bibinfo  {journal} {Phys. Rev. X}\ }\textbf {\bibinfo {volume} {9}},\
  \bibinfo {pages} {031040} (\bibinfo {year} {2019})},\ \Eprint
  {http://arxiv.org/abs/1811.12907} {arXiv:1811.12907 [astro-ph.HE]}
  \BibitemShut {NoStop}%
\bibitem [{\citenamefont {Abbott}\ \emph {et~al.}(2021)\citenamefont {Abbott}
  \emph {et~al.}}]{LIGO_GWTC2}%
  \BibitemOpen
  \bibfield  {author} {\bibinfo {author} {\bibfnamefont {R.}~\bibnamefont
  {Abbott}} \emph {et~al.} (\bibinfo {collaboration} {LIGO Scientific,
  Virgo}),\ }\href {\doibase 10.1103/PhysRevX.11.021053} {\bibfield  {journal}
  {\bibinfo  {journal} {Phys. Rev. X}\ }\textbf {\bibinfo {volume} {11}},\
  \bibinfo {pages} {021053} (\bibinfo {year} {2021})},\ \Eprint
  {http://arxiv.org/abs/2010.14527} {arXiv:2010.14527 [gr-qc]} \BibitemShut
  {NoStop}%
\bibitem [{\citenamefont {Barausse}\ \emph {et~al.}(2014)\citenamefont
  {Barausse}, \citenamefont {Cardoso},\ and\ \citenamefont
  {Pani}}]{Barausse:2014tra}%
  \BibitemOpen
  \bibfield  {author} {\bibinfo {author} {\bibfnamefont {E.}~\bibnamefont
  {Barausse}}, \bibinfo {author} {\bibfnamefont {V.}~\bibnamefont {Cardoso}}, \
  and\ \bibinfo {author} {\bibfnamefont {P.}~\bibnamefont {Pani}},\ }\href
  {\doibase 10.1103/PhysRevD.89.104059} {\bibfield  {journal} {\bibinfo
  {journal} {Phys. Rev. D}\ }\textbf {\bibinfo {volume} {89}},\ \bibinfo
  {pages} {104059} (\bibinfo {year} {2014})},\ \Eprint
  {http://arxiv.org/abs/1404.7149} {arXiv:1404.7149 [gr-qc]} \BibitemShut
  {NoStop}%
\bibitem [{\citenamefont {Cardoso}\ \emph
  {et~al.}(2016{\natexlab{a}})\citenamefont {Cardoso}, \citenamefont
  {Franzin},\ and\ \citenamefont {Pani}}]{Cardoso:2016rao}%
  \BibitemOpen
  \bibfield  {author} {\bibinfo {author} {\bibfnamefont {V.}~\bibnamefont
  {Cardoso}}, \bibinfo {author} {\bibfnamefont {E.}~\bibnamefont {Franzin}}, \
  and\ \bibinfo {author} {\bibfnamefont {P.}~\bibnamefont {Pani}},\ }\href
  {\doibase 10.1103/PhysRevLett.116.171101} {\bibfield  {journal} {\bibinfo
  {journal} {Phys. Rev. Lett.}\ }\textbf {\bibinfo {volume} {116}},\ \bibinfo
  {pages} {171101} (\bibinfo {year} {2016}{\natexlab{a}})},\ \bibinfo {note}
  {[Erratum: Phys.Rev.Lett. 117, 089902 (2016)]},\ \Eprint
  {http://arxiv.org/abs/1602.07309} {arXiv:1602.07309 [gr-qc]} \BibitemShut
  {NoStop}%
\bibitem [{\citenamefont {Unruh}\ and\ \citenamefont
  {Wald}(2017)}]{Unruh:2017uaw}%
  \BibitemOpen
  \bibfield  {author} {\bibinfo {author} {\bibfnamefont {W.~G.}\ \bibnamefont
  {Unruh}}\ and\ \bibinfo {author} {\bibfnamefont {R.~M.}\ \bibnamefont
  {Wald}},\ }\href {\doibase 10.1088/1361-6633/aa778e} {\bibfield  {journal}
  {\bibinfo  {journal} {Rept. Prog. Phys.}\ }\textbf {\bibinfo {volume} {80}},\
  \bibinfo {pages} {092002} (\bibinfo {year} {2017})},\ \Eprint
  {http://arxiv.org/abs/1703.02140} {arXiv:1703.02140 [hep-th]} \BibitemShut
  {NoStop}%
\bibitem [{\citenamefont {Mathur}(2005)}]{Mathur:2005zp}%
  \BibitemOpen
  \bibfield  {author} {\bibinfo {author} {\bibfnamefont {S.~D.}\ \bibnamefont
  {Mathur}},\ }\href {\doibase 10.1002/prop.200410203} {\bibfield  {journal}
  {\bibinfo  {journal} {Fortsch. Phys.}\ }\textbf {\bibinfo {volume} {53}},\
  \bibinfo {pages} {793} (\bibinfo {year} {2005})},\ \Eprint
  {http://arxiv.org/abs/hep-th/0502050} {arXiv:hep-th/0502050} \BibitemShut
  {NoStop}%
\bibitem [{\citenamefont {Mathur}(2008)}]{Mathur:2008nj}%
  \BibitemOpen
  \bibfield  {author} {\bibinfo {author} {\bibfnamefont {S.~D.}\ \bibnamefont
  {Mathur}},\ }\href@noop {} {\  (\bibinfo {year} {2008})},\ \Eprint
  {http://arxiv.org/abs/0810.4525} {arXiv:0810.4525 [hep-th]} \BibitemShut
  {NoStop}%
\bibitem [{\citenamefont {Almheiri}\ \emph {et~al.}(2013)\citenamefont
  {Almheiri}, \citenamefont {Marolf}, \citenamefont {Polchinski},\ and\
  \citenamefont {Sully}}]{Almheiri:2012rt}%
  \BibitemOpen
  \bibfield  {author} {\bibinfo {author} {\bibfnamefont {A.}~\bibnamefont
  {Almheiri}}, \bibinfo {author} {\bibfnamefont {D.}~\bibnamefont {Marolf}},
  \bibinfo {author} {\bibfnamefont {J.}~\bibnamefont {Polchinski}}, \ and\
  \bibinfo {author} {\bibfnamefont {J.}~\bibnamefont {Sully}},\ }\href
  {\doibase 10.1007/JHEP02(2013)062} {\bibfield  {journal} {\bibinfo  {journal}
  {JHEP}\ }\textbf {\bibinfo {volume} {02}},\ \bibinfo {pages} {062} (\bibinfo
  {year} {2013})},\ \Eprint {http://arxiv.org/abs/1207.3123} {arXiv:1207.3123
  [hep-th]} \BibitemShut {NoStop}%
\bibitem [{\citenamefont {Giddings}\ and\ \citenamefont
  {Psaltis}(2018)}]{Giddings_EHT2018}%
  \BibitemOpen
  \bibfield  {author} {\bibinfo {author} {\bibfnamefont {S.~B.}\ \bibnamefont
  {Giddings}}\ and\ \bibinfo {author} {\bibfnamefont {D.}~\bibnamefont
  {Psaltis}},\ }\href {\doibase 10.1103/PhysRevD.97.084035} {\bibfield
  {journal} {\bibinfo  {journal} {Phys. Rev. D}\ }\textbf {\bibinfo {volume}
  {97}},\ \bibinfo {pages} {084035} (\bibinfo {year} {2018})}\BibitemShut
  {NoStop}%
\bibitem [{\citenamefont {Bianchi}\ \emph {et~al.}(2020)\citenamefont
  {Bianchi}, \citenamefont {Consoli}, \citenamefont {Grillo}, \citenamefont
  {Morales}, \citenamefont {Pani},\ and\ \citenamefont
  {Raposo}}]{Bianchi_fuzzballs2020}%
  \BibitemOpen
  \bibfield  {author} {\bibinfo {author} {\bibfnamefont {M.}~\bibnamefont
  {Bianchi}}, \bibinfo {author} {\bibfnamefont {D.}~\bibnamefont {Consoli}},
  \bibinfo {author} {\bibfnamefont {A.}~\bibnamefont {Grillo}}, \bibinfo
  {author} {\bibfnamefont {J.~F.}\ \bibnamefont {Morales}}, \bibinfo {author}
  {\bibfnamefont {P.}~\bibnamefont {Pani}}, \ and\ \bibinfo {author}
  {\bibfnamefont {G.}~\bibnamefont {Raposo}},\ }\href {\doibase
  10.1103/PhysRevLett.125.221601} {\bibfield  {journal} {\bibinfo  {journal}
  {Phys. Rev. Lett.}\ }\textbf {\bibinfo {volume} {125}},\ \bibinfo {pages}
  {221601} (\bibinfo {year} {2020})},\ \Eprint
  {http://arxiv.org/abs/2007.01743} {arXiv:2007.01743 [hep-th]} \BibitemShut
  {NoStop}%
\bibitem [{\citenamefont {Bena}\ and\ \citenamefont
  {Mayerson}(2020)}]{Bena_MultiRatio2020}%
  \BibitemOpen
  \bibfield  {author} {\bibinfo {author} {\bibfnamefont {I.}~\bibnamefont
  {Bena}}\ and\ \bibinfo {author} {\bibfnamefont {D.~R.}\ \bibnamefont
  {Mayerson}},\ }\href {\doibase 10.1103/PhysRevLett.125.221602} {\bibfield
  {journal} {\bibinfo  {journal} {Phys. Rev. Lett.}\ }\textbf {\bibinfo
  {volume} {125}},\ \bibinfo {pages} {221602} (\bibinfo {year}
  {2020})}\BibitemShut {NoStop}%
\bibitem [{\citenamefont {Mukherjee}\ and\ \citenamefont
  {Chakraborty}(2020)}]{Mukherjee_Multimoments2020}%
  \BibitemOpen
  \bibfield  {author} {\bibinfo {author} {\bibfnamefont {S.}~\bibnamefont
  {Mukherjee}}\ and\ \bibinfo {author} {\bibfnamefont {S.}~\bibnamefont
  {Chakraborty}},\ }\href {\doibase 10.1103/PhysRevD.102.124058} {\bibfield
  {journal} {\bibinfo  {journal} {Phys. Rev. D}\ }\textbf {\bibinfo {volume}
  {102}},\ \bibinfo {pages} {124058} (\bibinfo {year} {2020})}\BibitemShut
  {NoStop}%
\bibitem [{\citenamefont {Buoninfante}\ \emph {et~al.}(2020)\citenamefont
  {Buoninfante}, \citenamefont {Lambiase}, \citenamefont {Luciano},\ and\
  \citenamefont {Petruzziello}}]{Buoninfante:2020cqz}%
  \BibitemOpen
  \bibfield  {author} {\bibinfo {author} {\bibfnamefont {L.}~\bibnamefont
  {Buoninfante}}, \bibinfo {author} {\bibfnamefont {G.}~\bibnamefont
  {Lambiase}}, \bibinfo {author} {\bibfnamefont {G.~G.}\ \bibnamefont
  {Luciano}}, \ and\ \bibinfo {author} {\bibfnamefont {L.}~\bibnamefont
  {Petruzziello}},\ }\href {\doibase 10.1140/epjc/s10052-020-08436-3}
  {\bibfield  {journal} {\bibinfo  {journal} {Eur. Phys. J. C}\ }\textbf
  {\bibinfo {volume} {80}},\ \bibinfo {pages} {853} (\bibinfo {year} {2020})},\
  \Eprint {http://arxiv.org/abs/2001.05825} {arXiv:2001.05825 [gr-qc]}
  \BibitemShut {NoStop}%
\bibitem [{\citenamefont {Mazur}\ and\ \citenamefont
  {Mottola}(2001)}]{Mazur:2001fv}%
  \BibitemOpen
  \bibfield  {author} {\bibinfo {author} {\bibfnamefont {P.~O.}\ \bibnamefont
  {Mazur}}\ and\ \bibinfo {author} {\bibfnamefont {E.}~\bibnamefont
  {Mottola}},\ }\href@noop {} {\  (\bibinfo {year} {2001})},\ \Eprint
  {http://arxiv.org/abs/gr-qc/0109035} {arXiv:gr-qc/0109035} \BibitemShut
  {NoStop}%
\bibitem [{\citenamefont {Mazur}\ and\ \citenamefont
  {Mottola}(2004)}]{Mazur_condensestars2004}%
  \BibitemOpen
  \bibfield  {author} {\bibinfo {author} {\bibfnamefont {P.~O.}\ \bibnamefont
  {Mazur}}\ and\ \bibinfo {author} {\bibfnamefont {E.}~\bibnamefont
  {Mottola}},\ }\href {\doibase 10.1073/pnas.0402717101} {\bibfield  {journal}
  {\bibinfo  {journal} {Proceedings of the National Academy of Sciences}\
  }\textbf {\bibinfo {volume} {101}},\ \bibinfo {pages} {9545} (\bibinfo {year}
  {2004})},\ \Eprint
  {http://arxiv.org/abs/https://www.pnas.org/content/101/26/9545.full.pdf}
  {https://www.pnas.org/content/101/26/9545.full.pdf} \BibitemShut {NoStop}%
\bibitem [{\citenamefont {Pani}\ \emph {et~al.}(2009)\citenamefont {Pani},
  \citenamefont {Berti}, \citenamefont {Cardoso}, \citenamefont {Chen},\ and\
  \citenamefont {Norte}}]{Pani_gravastar2009}%
  \BibitemOpen
  \bibfield  {author} {\bibinfo {author} {\bibfnamefont {P.}~\bibnamefont
  {Pani}}, \bibinfo {author} {\bibfnamefont {E.}~\bibnamefont {Berti}},
  \bibinfo {author} {\bibfnamefont {V.}~\bibnamefont {Cardoso}}, \bibinfo
  {author} {\bibfnamefont {Y.}~\bibnamefont {Chen}}, \ and\ \bibinfo {author}
  {\bibfnamefont {R.}~\bibnamefont {Norte}},\ }\href {\doibase
  10.1103/PhysRevD.80.124047} {\bibfield  {journal} {\bibinfo  {journal} {Phys.
  Rev. D}\ }\textbf {\bibinfo {volume} {80}},\ \bibinfo {pages} {124047}
  (\bibinfo {year} {2009})}\BibitemShut {NoStop}%
\bibitem [{\citenamefont {Palenzuela}\ \emph {et~al.}(2017)\citenamefont
  {Palenzuela}, \citenamefont {Pani}, \citenamefont {Bezares}, \citenamefont
  {Cardoso}, \citenamefont {Lehner},\ and\ \citenamefont
  {Liebling}}]{Carlos_bosonstar2017}%
  \BibitemOpen
  \bibfield  {author} {\bibinfo {author} {\bibfnamefont {C.}~\bibnamefont
  {Palenzuela}}, \bibinfo {author} {\bibfnamefont {P.}~\bibnamefont {Pani}},
  \bibinfo {author} {\bibfnamefont {M.}~\bibnamefont {Bezares}}, \bibinfo
  {author} {\bibfnamefont {V.}~\bibnamefont {Cardoso}}, \bibinfo {author}
  {\bibfnamefont {L.}~\bibnamefont {Lehner}}, \ and\ \bibinfo {author}
  {\bibfnamefont {S.}~\bibnamefont {Liebling}},\ }\href {\doibase
  10.1103/PhysRevD.96.104058} {\bibfield  {journal} {\bibinfo  {journal} {Phys.
  Rev. D}\ }\textbf {\bibinfo {volume} {96}},\ \bibinfo {pages} {104058}
  (\bibinfo {year} {2017})}\BibitemShut {NoStop}%
\bibitem [{\citenamefont {Cardoso}\ and\ \citenamefont
  {Pani}(2019{\natexlab{a}})}]{Cardoso:2019rvt}%
  \BibitemOpen
  \bibfield  {author} {\bibinfo {author} {\bibfnamefont {V.}~\bibnamefont
  {Cardoso}}\ and\ \bibinfo {author} {\bibfnamefont {P.}~\bibnamefont {Pani}},\
  }\href {\doibase 10.1007/s41114-019-0020-4} {\bibfield  {journal} {\bibinfo
  {journal} {Living Rev. Rel.}\ }\textbf {\bibinfo {volume} {22}},\ \bibinfo
  {pages} {4} (\bibinfo {year} {2019}{\natexlab{a}})},\ \Eprint
  {http://arxiv.org/abs/1904.05363} {arXiv:1904.05363 [gr-qc]} \BibitemShut
  {NoStop}%
\bibitem [{\citenamefont {Hughes}(2013)}]{Hughes2013}%
  \BibitemOpen
  \bibfield  {author} {\bibinfo {author} {\bibfnamefont {S.~A.}\ \bibnamefont
  {Hughes}},\ }\href {\doibase 10.1103/PhysRevD.88.109902} {\bibfield
  {journal} {\bibinfo  {journal} {Phys. Rev. D}\ }\textbf {\bibinfo {volume}
  {88}},\ \bibinfo {pages} {109902} (\bibinfo {year} {2013})}\BibitemShut
  {NoStop}%
\bibitem [{\citenamefont {Krishnendu}\ \emph {et~al.}(2017)\citenamefont
  {Krishnendu}, \citenamefont {Arun},\ and\ \citenamefont
  {Mishra}}]{Krishnendu_2017}%
  \BibitemOpen
  \bibfield  {author} {\bibinfo {author} {\bibfnamefont {N.~V.}\ \bibnamefont
  {Krishnendu}}, \bibinfo {author} {\bibfnamefont {K.~G.}\ \bibnamefont
  {Arun}}, \ and\ \bibinfo {author} {\bibfnamefont {C.~K.}\ \bibnamefont
  {Mishra}},\ }\href {\doibase 10.1103/PhysRevLett.119.091101} {\bibfield
  {journal} {\bibinfo  {journal} {Phys. Rev. Lett.}\ }\textbf {\bibinfo
  {volume} {119}},\ \bibinfo {pages} {091101} (\bibinfo {year}
  {2017})}\BibitemShut {NoStop}%
\bibitem [{\citenamefont {Cardoso}\ \emph {et~al.}(2017)\citenamefont
  {Cardoso}, \citenamefont {Franzin}, \citenamefont {Maselli}, \citenamefont
  {Pani},\ and\ \citenamefont {Raposo}}]{Cardoso_tidal2017}%
  \BibitemOpen
  \bibfield  {author} {\bibinfo {author} {\bibfnamefont {V.}~\bibnamefont
  {Cardoso}}, \bibinfo {author} {\bibfnamefont {E.}~\bibnamefont {Franzin}},
  \bibinfo {author} {\bibfnamefont {A.}~\bibnamefont {Maselli}}, \bibinfo
  {author} {\bibfnamefont {P.}~\bibnamefont {Pani}}, \ and\ \bibinfo {author}
  {\bibfnamefont {G.}~\bibnamefont {Raposo}},\ }\href {\doibase
  10.1103/PhysRevD.95.084014} {\bibfield  {journal} {\bibinfo  {journal} {Phys.
  Rev. D}\ }\textbf {\bibinfo {volume} {95}},\ \bibinfo {pages} {084014}
  (\bibinfo {year} {2017})}\BibitemShut {NoStop}%
\bibitem [{\citenamefont {Maselli}\ \emph {et~al.}(2018)\citenamefont
  {Maselli}, \citenamefont {Pani}, \citenamefont {Cardoso}, \citenamefont
  {Abdelsalhin}, \citenamefont {Gualtieri},\ and\ \citenamefont
  {Ferrari}}]{Maselli_Planckcorrect2018}%
  \BibitemOpen
  \bibfield  {author} {\bibinfo {author} {\bibfnamefont {A.}~\bibnamefont
  {Maselli}}, \bibinfo {author} {\bibfnamefont {P.}~\bibnamefont {Pani}},
  \bibinfo {author} {\bibfnamefont {V.}~\bibnamefont {Cardoso}}, \bibinfo
  {author} {\bibfnamefont {T.}~\bibnamefont {Abdelsalhin}}, \bibinfo {author}
  {\bibfnamefont {L.}~\bibnamefont {Gualtieri}}, \ and\ \bibinfo {author}
  {\bibfnamefont {V.}~\bibnamefont {Ferrari}},\ }\href {\doibase
  10.1103/PhysRevLett.120.081101} {\bibfield  {journal} {\bibinfo  {journal}
  {Phys. Rev. Lett.}\ }\textbf {\bibinfo {volume} {120}},\ \bibinfo {pages}
  {081101} (\bibinfo {year} {2018})}\BibitemShut {NoStop}%
\bibitem [{\citenamefont {Sennett}\ \emph {et~al.}(2017)\citenamefont
  {Sennett}, \citenamefont {Hinderer}, \citenamefont {Steinhoff}, \citenamefont
  {Buonanno},\ and\ \citenamefont {Ossokine}}]{Sennett_tidal2017}%
  \BibitemOpen
  \bibfield  {author} {\bibinfo {author} {\bibfnamefont {N.}~\bibnamefont
  {Sennett}}, \bibinfo {author} {\bibfnamefont {T.}~\bibnamefont {Hinderer}},
  \bibinfo {author} {\bibfnamefont {J.}~\bibnamefont {Steinhoff}}, \bibinfo
  {author} {\bibfnamefont {A.}~\bibnamefont {Buonanno}}, \ and\ \bibinfo
  {author} {\bibfnamefont {S.}~\bibnamefont {Ossokine}},\ }\href {\doibase
  10.1103/PhysRevD.96.024002} {\bibfield  {journal} {\bibinfo  {journal} {Phys.
  Rev. D}\ }\textbf {\bibinfo {volume} {96}},\ \bibinfo {pages} {024002}
  (\bibinfo {year} {2017})}\BibitemShut {NoStop}%
\bibitem [{\citenamefont {Krishnendu}\ \emph {et~al.}(2019)\citenamefont
  {Krishnendu}, \citenamefont {Mishra},\ and\ \citenamefont
  {Arun}}]{Krishnendu2019}%
  \BibitemOpen
  \bibfield  {author} {\bibinfo {author} {\bibfnamefont {N.~V.}\ \bibnamefont
  {Krishnendu}}, \bibinfo {author} {\bibfnamefont {C.~K.}\ \bibnamefont
  {Mishra}}, \ and\ \bibinfo {author} {\bibfnamefont {K.~G.}\ \bibnamefont
  {Arun}},\ }\href {\doibase 10.1103/PhysRevD.99.064008} {\bibfield  {journal}
  {\bibinfo  {journal} {Phys. Rev. D}\ }\textbf {\bibinfo {volume} {99}},\
  \bibinfo {pages} {064008} (\bibinfo {year} {2019})}\BibitemShut {NoStop}%
\bibitem [{\citenamefont {Datta}\ and\ \citenamefont {Bose}(2019)}]{Datta2019}%
  \BibitemOpen
  \bibfield  {author} {\bibinfo {author} {\bibfnamefont {S.}~\bibnamefont
  {Datta}}\ and\ \bibinfo {author} {\bibfnamefont {S.}~\bibnamefont {Bose}},\
  }\href {\doibase 10.1103/PhysRevD.99.084001} {\bibfield  {journal} {\bibinfo
  {journal} {Phys. Rev. D}\ }\textbf {\bibinfo {volume} {99}},\ \bibinfo
  {pages} {084001} (\bibinfo {year} {2019})}\BibitemShut {NoStop}%
\bibitem [{\citenamefont {Datta}\ \emph {et~al.}(2020)\citenamefont {Datta},
  \citenamefont {Brito}, \citenamefont {Bose}, \citenamefont {Pani},\ and\
  \citenamefont {Hughes}}]{Datta_tidal2019}%
  \BibitemOpen
  \bibfield  {author} {\bibinfo {author} {\bibfnamefont {S.}~\bibnamefont
  {Datta}}, \bibinfo {author} {\bibfnamefont {R.}~\bibnamefont {Brito}},
  \bibinfo {author} {\bibfnamefont {S.}~\bibnamefont {Bose}}, \bibinfo {author}
  {\bibfnamefont {P.}~\bibnamefont {Pani}}, \ and\ \bibinfo {author}
  {\bibfnamefont {S.~A.}\ \bibnamefont {Hughes}},\ }\href {\doibase
  10.1103/PhysRevD.101.044004} {\bibfield  {journal} {\bibinfo  {journal}
  {Phys. Rev. D}\ }\textbf {\bibinfo {volume} {101}},\ \bibinfo {pages}
  {044004} (\bibinfo {year} {2020})},\ \Eprint
  {http://arxiv.org/abs/1910.07841} {arXiv:1910.07841 [gr-qc]} \BibitemShut
  {NoStop}%
\bibitem [{\citenamefont {Dreyer}\ \emph {et~al.}(2004)\citenamefont {Dreyer},
  \citenamefont {Kelly}, \citenamefont {Krishnan}, \citenamefont {Finn},
  \citenamefont {Garrison},\ and\ \citenamefont
  {Lopez-Aleman}}]{Dreyer:2003bv}%
  \BibitemOpen
  \bibfield  {author} {\bibinfo {author} {\bibfnamefont {O.}~\bibnamefont
  {Dreyer}}, \bibinfo {author} {\bibfnamefont {B.~J.}\ \bibnamefont {Kelly}},
  \bibinfo {author} {\bibfnamefont {B.}~\bibnamefont {Krishnan}}, \bibinfo
  {author} {\bibfnamefont {L.~S.}\ \bibnamefont {Finn}}, \bibinfo {author}
  {\bibfnamefont {D.}~\bibnamefont {Garrison}}, \ and\ \bibinfo {author}
  {\bibfnamefont {R.}~\bibnamefont {Lopez-Aleman}},\ }\href {\doibase
  10.1088/0264-9381/21/4/003} {\bibfield  {journal} {\bibinfo  {journal}
  {Class. Quant. Grav.}\ }\textbf {\bibinfo {volume} {21}},\ \bibinfo {pages}
  {787} (\bibinfo {year} {2004})},\ \Eprint
  {http://arxiv.org/abs/gr-qc/0309007} {arXiv:gr-qc/0309007} \BibitemShut
  {NoStop}%
\bibitem [{\citenamefont {Berti}\ \emph {et~al.}(2006)\citenamefont {Berti},
  \citenamefont {Cardoso},\ and\ \citenamefont {Will}}]{Berti:2005ys}%
  \BibitemOpen
  \bibfield  {author} {\bibinfo {author} {\bibfnamefont {E.}~\bibnamefont
  {Berti}}, \bibinfo {author} {\bibfnamefont {V.}~\bibnamefont {Cardoso}}, \
  and\ \bibinfo {author} {\bibfnamefont {C.~M.}\ \bibnamefont {Will}},\ }\href
  {\doibase 10.1103/PhysRevD.73.064030} {\bibfield  {journal} {\bibinfo
  {journal} {Phys. Rev. D}\ }\textbf {\bibinfo {volume} {73}},\ \bibinfo
  {pages} {064030} (\bibinfo {year} {2006})},\ \Eprint
  {http://arxiv.org/abs/gr-qc/0512160} {arXiv:gr-qc/0512160} \BibitemShut
  {NoStop}%
\bibitem [{\citenamefont {Berti}\ \emph {et~al.}(2016)\citenamefont {Berti},
  \citenamefont {Sesana}, \citenamefont {Barausse}, \citenamefont {Cardoso},\
  and\ \citenamefont {Belczynski}}]{Berti:2016lat}%
  \BibitemOpen
  \bibfield  {author} {\bibinfo {author} {\bibfnamefont {E.}~\bibnamefont
  {Berti}}, \bibinfo {author} {\bibfnamefont {A.}~\bibnamefont {Sesana}},
  \bibinfo {author} {\bibfnamefont {E.}~\bibnamefont {Barausse}}, \bibinfo
  {author} {\bibfnamefont {V.}~\bibnamefont {Cardoso}}, \ and\ \bibinfo
  {author} {\bibfnamefont {K.}~\bibnamefont {Belczynski}},\ }\href {\doibase
  10.1103/PhysRevLett.117.101102} {\bibfield  {journal} {\bibinfo  {journal}
  {Phys. Rev. Lett.}\ }\textbf {\bibinfo {volume} {117}},\ \bibinfo {pages}
  {101102} (\bibinfo {year} {2016})},\ \Eprint
  {http://arxiv.org/abs/1605.09286} {arXiv:1605.09286 [gr-qc]} \BibitemShut
  {NoStop}%
\bibitem [{\citenamefont {Berti}\ and\ \citenamefont
  {Cardoso}(2006)}]{Berti:2006qt}%
  \BibitemOpen
  \bibfield  {author} {\bibinfo {author} {\bibfnamefont {E.}~\bibnamefont
  {Berti}}\ and\ \bibinfo {author} {\bibfnamefont {V.}~\bibnamefont
  {Cardoso}},\ }\href {\doibase 10.1142/S0218271806009637} {\bibfield
  {journal} {\bibinfo  {journal} {Int. J. Mod. Phys. D}\ }\textbf {\bibinfo
  {volume} {15}},\ \bibinfo {pages} {2209} (\bibinfo {year} {2006})},\ \Eprint
  {http://arxiv.org/abs/gr-qc/0605101} {arXiv:gr-qc/0605101} \BibitemShut
  {NoStop}%
\bibitem [{\citenamefont {Cardoso}\ \emph {et~al.}(2019)\citenamefont
  {Cardoso}, \citenamefont {del Rio},\ and\ \citenamefont
  {Kimura}}]{Cardoso_EMRI_ECO2019}%
  \BibitemOpen
  \bibfield  {author} {\bibinfo {author} {\bibfnamefont {V.}~\bibnamefont
  {Cardoso}}, \bibinfo {author} {\bibfnamefont {A.}~\bibnamefont {del Rio}}, \
  and\ \bibinfo {author} {\bibfnamefont {M.}~\bibnamefont {Kimura}},\ }\href
  {\doibase 10.1103/PhysRevD.100.084046} {\bibfield  {journal} {\bibinfo
  {journal} {Phys. Rev. D}\ }\textbf {\bibinfo {volume} {100}},\ \bibinfo
  {pages} {084046} (\bibinfo {year} {2019})}\BibitemShut {NoStop}%
\bibitem [{\citenamefont {Maggio}\ \emph
  {et~al.}(2021{\natexlab{a}})\citenamefont {Maggio}, \citenamefont {van~de
  Meent},\ and\ \citenamefont {Pani}}]{Maggio_EMRI_ECO2021}%
  \BibitemOpen
  \bibfield  {author} {\bibinfo {author} {\bibfnamefont {E.}~\bibnamefont
  {Maggio}}, \bibinfo {author} {\bibfnamefont {M.}~\bibnamefont {van~de
  Meent}}, \ and\ \bibinfo {author} {\bibfnamefont {P.}~\bibnamefont {Pani}},\
  }\href@noop {} {\  (\bibinfo {year} {2021}{\natexlab{a}})},\ \Eprint
  {http://arxiv.org/abs/2106.07195} {arXiv:2106.07195 [gr-qc]} \BibitemShut
  {NoStop}%
\bibitem [{\citenamefont {Sago}\ and\ \citenamefont
  {Tanaka}(2021)}]{Sago_EMRI2021}%
  \BibitemOpen
  \bibfield  {author} {\bibinfo {author} {\bibfnamefont {N.}~\bibnamefont
  {Sago}}\ and\ \bibinfo {author} {\bibfnamefont {T.}~\bibnamefont {Tanaka}},\
  }\href@noop {} {\  (\bibinfo {year} {2021})},\ \Eprint
  {http://arxiv.org/abs/2106.07123} {arXiv:2106.07123 [gr-qc]} \BibitemShut
  {NoStop}%
\bibitem [{\citenamefont {Cardoso}\ \emph
  {et~al.}(2016{\natexlab{b}})\citenamefont {Cardoso}, \citenamefont
  {Franzin},\ and\ \citenamefont {Pani}}]{Cardoso_echo1st2016}%
  \BibitemOpen
  \bibfield  {author} {\bibinfo {author} {\bibfnamefont {V.}~\bibnamefont
  {Cardoso}}, \bibinfo {author} {\bibfnamefont {E.}~\bibnamefont {Franzin}}, \
  and\ \bibinfo {author} {\bibfnamefont {P.}~\bibnamefont {Pani}},\ }\href
  {\doibase 10.1103/PhysRevLett.117.089902} {\bibfield  {journal} {\bibinfo
  {journal} {Phys. Rev. Lett.}\ }\textbf {\bibinfo {volume} {117}},\ \bibinfo
  {pages} {089902} (\bibinfo {year} {2016}{\natexlab{b}})}\BibitemShut
  {NoStop}%
\bibitem [{\citenamefont {Cardoso}\ \emph
  {et~al.}(2016{\natexlab{c}})\citenamefont {Cardoso}, \citenamefont {Hopper},
  \citenamefont {Macedo}, \citenamefont {Palenzuela},\ and\ \citenamefont
  {Pani}}]{Cardoso_quantumEH2016}%
  \BibitemOpen
  \bibfield  {author} {\bibinfo {author} {\bibfnamefont {V.}~\bibnamefont
  {Cardoso}}, \bibinfo {author} {\bibfnamefont {S.}~\bibnamefont {Hopper}},
  \bibinfo {author} {\bibfnamefont {C.~F.~B.}\ \bibnamefont {Macedo}}, \bibinfo
  {author} {\bibfnamefont {C.}~\bibnamefont {Palenzuela}}, \ and\ \bibinfo
  {author} {\bibfnamefont {P.}~\bibnamefont {Pani}},\ }\href {\doibase
  10.1103/PhysRevD.94.084031} {\bibfield  {journal} {\bibinfo  {journal} {Phys.
  Rev. D}\ }\textbf {\bibinfo {volume} {94}},\ \bibinfo {pages} {084031}
  (\bibinfo {year} {2016}{\natexlab{c}})}\BibitemShut {NoStop}%
\bibitem [{\citenamefont {{Cardoso}}\ and\ \citenamefont
  {{Pani}}(2017)}]{Cardoso_2017natureAstro}%
  \BibitemOpen
  \bibfield  {author} {\bibinfo {author} {\bibfnamefont {V.}~\bibnamefont
  {{Cardoso}}}\ and\ \bibinfo {author} {\bibfnamefont {P.}~\bibnamefont
  {{Pani}}},\ }\href {\doibase 10.1038/s41550-017-0225-y} {\bibfield  {journal}
  {\bibinfo  {journal} {Nature Astronomy}\ }\textbf {\bibinfo {volume} {1}},\
  \bibinfo {pages} {586} (\bibinfo {year} {2017})},\ \Eprint
  {http://arxiv.org/abs/1707.03021} {arXiv:1707.03021 [gr-qc]} \BibitemShut
  {NoStop}%
\bibitem [{\citenamefont {Abedi}\ \emph {et~al.}(2017)\citenamefont {Abedi},
  \citenamefont {Dykaar},\ and\ \citenamefont {Afshordi}}]{Abedi2017}%
  \BibitemOpen
  \bibfield  {author} {\bibinfo {author} {\bibfnamefont {J.}~\bibnamefont
  {Abedi}}, \bibinfo {author} {\bibfnamefont {H.}~\bibnamefont {Dykaar}}, \
  and\ \bibinfo {author} {\bibfnamefont {N.}~\bibnamefont {Afshordi}},\ }\href
  {\doibase 10.1103/PhysRevD.96.082004} {\bibfield  {journal} {\bibinfo
  {journal} {Phys. Rev. D}\ }\textbf {\bibinfo {volume} {96}},\ \bibinfo
  {pages} {082004} (\bibinfo {year} {2017})}\BibitemShut {NoStop}%
\bibitem [{\citenamefont {Conklin}\ \emph {et~al.}(2018)\citenamefont
  {Conklin}, \citenamefont {Holdom},\ and\ \citenamefont {Ren}}]{Conklin2018}%
  \BibitemOpen
  \bibfield  {author} {\bibinfo {author} {\bibfnamefont {R.~S.}\ \bibnamefont
  {Conklin}}, \bibinfo {author} {\bibfnamefont {B.}~\bibnamefont {Holdom}}, \
  and\ \bibinfo {author} {\bibfnamefont {J.}~\bibnamefont {Ren}},\ }\href
  {\doibase 10.1103/PhysRevD.98.044021} {\bibfield  {journal} {\bibinfo
  {journal} {Phys. Rev. D}\ }\textbf {\bibinfo {volume} {98}},\ \bibinfo
  {pages} {044021} (\bibinfo {year} {2018})}\BibitemShut {NoStop}%
\bibitem [{\citenamefont {Tsang}\ \emph {et~al.}(2018)\citenamefont {Tsang},
  \citenamefont {Rollier}, \citenamefont {Ghosh}, \citenamefont {Samajdar},
  \citenamefont {Agathos}, \citenamefont {Chatziioannou}, \citenamefont
  {Cardoso}, \citenamefont {Khanna},\ and\ \citenamefont {Van
  Den~Broeck}}]{Tsang2018}%
  \BibitemOpen
  \bibfield  {author} {\bibinfo {author} {\bibfnamefont {K.~W.}\ \bibnamefont
  {Tsang}}, \bibinfo {author} {\bibfnamefont {M.}~\bibnamefont {Rollier}},
  \bibinfo {author} {\bibfnamefont {A.}~\bibnamefont {Ghosh}}, \bibinfo
  {author} {\bibfnamefont {A.}~\bibnamefont {Samajdar}}, \bibinfo {author}
  {\bibfnamefont {M.}~\bibnamefont {Agathos}}, \bibinfo {author} {\bibfnamefont
  {K.}~\bibnamefont {Chatziioannou}}, \bibinfo {author} {\bibfnamefont
  {V.}~\bibnamefont {Cardoso}}, \bibinfo {author} {\bibfnamefont
  {G.}~\bibnamefont {Khanna}}, \ and\ \bibinfo {author} {\bibfnamefont
  {C.}~\bibnamefont {Van Den~Broeck}},\ }\href {\doibase
  10.1103/PhysRevD.98.024023} {\bibfield  {journal} {\bibinfo  {journal} {Phys.
  Rev. D}\ }\textbf {\bibinfo {volume} {98}},\ \bibinfo {pages} {024023}
  (\bibinfo {year} {2018})}\BibitemShut {NoStop}%
\bibitem [{\citenamefont {Lo}\ \emph {et~al.}(2019)\citenamefont {Lo},
  \citenamefont {Li},\ and\ \citenamefont {Weinstein}}]{Rico2019}%
  \BibitemOpen
  \bibfield  {author} {\bibinfo {author} {\bibfnamefont {R.~K.~L.}\
  \bibnamefont {Lo}}, \bibinfo {author} {\bibfnamefont {T.~G.~F.}\ \bibnamefont
  {Li}}, \ and\ \bibinfo {author} {\bibfnamefont {A.~J.}\ \bibnamefont
  {Weinstein}},\ }\href {\doibase 10.1103/PhysRevD.99.084052} {\bibfield
  {journal} {\bibinfo  {journal} {Phys. Rev. D}\ }\textbf {\bibinfo {volume}
  {99}},\ \bibinfo {pages} {084052} (\bibinfo {year} {2019})}\BibitemShut
  {NoStop}%
\bibitem [{\citenamefont {Nielsen}\ \emph {et~al.}(2019)\citenamefont
  {Nielsen}, \citenamefont {Capano}, \citenamefont {Birnholtz},\ and\
  \citenamefont {Westerweck}}]{Nielsen_2019}%
  \BibitemOpen
  \bibfield  {author} {\bibinfo {author} {\bibfnamefont {A.~B.}\ \bibnamefont
  {Nielsen}}, \bibinfo {author} {\bibfnamefont {C.~D.}\ \bibnamefont {Capano}},
  \bibinfo {author} {\bibfnamefont {O.}~\bibnamefont {Birnholtz}}, \ and\
  \bibinfo {author} {\bibfnamefont {J.}~\bibnamefont {Westerweck}},\ }\href
  {\doibase 10.1103/PhysRevD.99.104012} {\bibfield  {journal} {\bibinfo
  {journal} {Phys. Rev. D}\ }\textbf {\bibinfo {volume} {99}},\ \bibinfo
  {pages} {104012} (\bibinfo {year} {2019})}\BibitemShut {NoStop}%
\bibitem [{\citenamefont {Uchikata}\ \emph {et~al.}(2019)\citenamefont
  {Uchikata}, \citenamefont {Nakano}, \citenamefont {Narikawa}, \citenamefont
  {Sago}, \citenamefont {Tagoshi},\ and\ \citenamefont
  {Tanaka}}]{Uchikata2019}%
  \BibitemOpen
  \bibfield  {author} {\bibinfo {author} {\bibfnamefont {N.}~\bibnamefont
  {Uchikata}}, \bibinfo {author} {\bibfnamefont {H.}~\bibnamefont {Nakano}},
  \bibinfo {author} {\bibfnamefont {T.}~\bibnamefont {Narikawa}}, \bibinfo
  {author} {\bibfnamefont {N.}~\bibnamefont {Sago}}, \bibinfo {author}
  {\bibfnamefont {H.}~\bibnamefont {Tagoshi}}, \ and\ \bibinfo {author}
  {\bibfnamefont {T.}~\bibnamefont {Tanaka}},\ }\href {\doibase
  10.1103/PhysRevD.100.062006} {\bibfield  {journal} {\bibinfo  {journal}
  {Phys. Rev. D}\ }\textbf {\bibinfo {volume} {100}},\ \bibinfo {pages}
  {062006} (\bibinfo {year} {2019})}\BibitemShut {NoStop}%
\bibitem [{\citenamefont {Tsang}\ \emph {et~al.}(2020)\citenamefont {Tsang},
  \citenamefont {Ghosh}, \citenamefont {Samajdar}, \citenamefont
  {Chatziioannou}, \citenamefont {Mastrogiovanni}, \citenamefont {Agathos},\
  and\ \citenamefont {Van Den~Broeck}}]{Tsang2020}%
  \BibitemOpen
  \bibfield  {author} {\bibinfo {author} {\bibfnamefont {K.~W.}\ \bibnamefont
  {Tsang}}, \bibinfo {author} {\bibfnamefont {A.}~\bibnamefont {Ghosh}},
  \bibinfo {author} {\bibfnamefont {A.}~\bibnamefont {Samajdar}}, \bibinfo
  {author} {\bibfnamefont {K.}~\bibnamefont {Chatziioannou}}, \bibinfo {author}
  {\bibfnamefont {S.}~\bibnamefont {Mastrogiovanni}}, \bibinfo {author}
  {\bibfnamefont {M.}~\bibnamefont {Agathos}}, \ and\ \bibinfo {author}
  {\bibfnamefont {C.}~\bibnamefont {Van Den~Broeck}},\ }\href {\doibase
  10.1103/PhysRevD.101.064012} {\bibfield  {journal} {\bibinfo  {journal}
  {Phys. Rev. D}\ }\textbf {\bibinfo {volume} {101}},\ \bibinfo {pages}
  {064012} (\bibinfo {year} {2020})},\ \Eprint
  {http://arxiv.org/abs/1906.11168} {arXiv:1906.11168 [gr-qc]} \BibitemShut
  {NoStop}%
\bibitem [{\citenamefont {Xin}\ \emph {et~al.}(2021)\citenamefont {Xin},
  \citenamefont {Chen}, \citenamefont {Lo}, \citenamefont {Sun}, \citenamefont
  {Han}, \citenamefont {Zhong}, \citenamefont {Srivastava}, \citenamefont {Ma},
  \citenamefont {Wang},\ and\ \citenamefont {Chen}}]{Xinshuo_echo2021}%
  \BibitemOpen
  \bibfield  {author} {\bibinfo {author} {\bibfnamefont {S.}~\bibnamefont
  {Xin}}, \bibinfo {author} {\bibfnamefont {B.}~\bibnamefont {Chen}}, \bibinfo
  {author} {\bibfnamefont {R.~K.~L.}\ \bibnamefont {Lo}}, \bibinfo {author}
  {\bibfnamefont {L.}~\bibnamefont {Sun}}, \bibinfo {author} {\bibfnamefont
  {W.-B.}\ \bibnamefont {Han}}, \bibinfo {author} {\bibfnamefont
  {X.}~\bibnamefont {Zhong}}, \bibinfo {author} {\bibfnamefont
  {M.}~\bibnamefont {Srivastava}}, \bibinfo {author} {\bibfnamefont
  {S.}~\bibnamefont {Ma}}, \bibinfo {author} {\bibfnamefont {Q.}~\bibnamefont
  {Wang}}, \ and\ \bibinfo {author} {\bibfnamefont {Y.}~\bibnamefont {Chen}},\
  }\href@noop {} {\  (\bibinfo {year} {2021})},\ \Eprint
  {http://arxiv.org/abs/2105.12313} {arXiv:2105.12313 [gr-qc]} \BibitemShut
  {NoStop}%
\bibitem [{\citenamefont {Barausse}\ \emph {et~al.}(2018)\citenamefont
  {Barausse}, \citenamefont {Brito}, \citenamefont {Cardoso}, \citenamefont
  {Dvorkin},\ and\ \citenamefont {Pani}}]{Barausse:2018vdb}%
  \BibitemOpen
  \bibfield  {author} {\bibinfo {author} {\bibfnamefont {E.}~\bibnamefont
  {Barausse}}, \bibinfo {author} {\bibfnamefont {R.}~\bibnamefont {Brito}},
  \bibinfo {author} {\bibfnamefont {V.}~\bibnamefont {Cardoso}}, \bibinfo
  {author} {\bibfnamefont {I.}~\bibnamefont {Dvorkin}}, \ and\ \bibinfo
  {author} {\bibfnamefont {P.}~\bibnamefont {Pani}},\ }\href {\doibase
  10.1088/1361-6382/aae1de} {\bibfield  {journal} {\bibinfo  {journal} {Class.
  Quant. Grav.}\ }\textbf {\bibinfo {volume} {35}},\ \bibinfo {pages} {20LT01}
  (\bibinfo {year} {2018})},\ \Eprint {http://arxiv.org/abs/1805.08229}
  {arXiv:1805.08229 [gr-qc]} \BibitemShut {NoStop}%
\bibitem [{\citenamefont {Fan}\ and\ \citenamefont {Chen}(2018)}]{Fan:2017cfw}%
  \BibitemOpen
  \bibfield  {author} {\bibinfo {author} {\bibfnamefont {X.-L.}\ \bibnamefont
  {Fan}}\ and\ \bibinfo {author} {\bibfnamefont {Y.-B.}\ \bibnamefont {Chen}},\
  }\href {\doibase 10.1103/PhysRevD.98.044020} {\bibfield  {journal} {\bibinfo
  {journal} {Phys. Rev. D}\ }\textbf {\bibinfo {volume} {98}},\ \bibinfo
  {pages} {044020} (\bibinfo {year} {2018})},\ \Eprint
  {http://arxiv.org/abs/1712.00784} {arXiv:1712.00784 [gr-qc]} \BibitemShut
  {NoStop}%
\bibitem [{\citenamefont {Du}\ and\ \citenamefont {Chen}(2018)}]{Du:2018cmp}%
  \BibitemOpen
  \bibfield  {author} {\bibinfo {author} {\bibfnamefont {S.~M.}\ \bibnamefont
  {Du}}\ and\ \bibinfo {author} {\bibfnamefont {Y.}~\bibnamefont {Chen}},\
  }\href {\doibase 10.1103/PhysRevLett.121.051105} {\bibfield  {journal}
  {\bibinfo  {journal} {Phys. Rev. Lett.}\ }\textbf {\bibinfo {volume} {121}},\
  \bibinfo {pages} {051105} (\bibinfo {year} {2018})},\ \Eprint
  {http://arxiv.org/abs/1803.10947} {arXiv:1803.10947 [gr-qc]} \BibitemShut
  {NoStop}%
\bibitem [{\citenamefont {Cardoso}\ and\ \citenamefont
  {Pani}(2019{\natexlab{b}})}]{Cardoso_review2019}%
  \BibitemOpen
  \bibfield  {author} {\bibinfo {author} {\bibfnamefont {V.}~\bibnamefont
  {Cardoso}}\ and\ \bibinfo {author} {\bibfnamefont {P.}~\bibnamefont {Pani}},\
  }\href {\doibase 10.1007/s41114-019-0020-4} {\bibfield  {journal} {\bibinfo
  {journal} {Living Rev. Rel.}\ }\textbf {\bibinfo {volume} {22}},\ \bibinfo
  {pages} {4} (\bibinfo {year} {2019}{\natexlab{b}})},\ \Eprint
  {http://arxiv.org/abs/1904.05363} {arXiv:1904.05363 [gr-qc]} \BibitemShut
  {NoStop}%
\bibitem [{\citenamefont {Maggio}\ \emph
  {et~al.}(2021{\natexlab{b}})\citenamefont {Maggio}, \citenamefont {Pani},\
  and\ \citenamefont {Raposo}}]{Maggio_review2021}%
  \BibitemOpen
  \bibfield  {author} {\bibinfo {author} {\bibfnamefont {E.}~\bibnamefont
  {Maggio}}, \bibinfo {author} {\bibfnamefont {P.}~\bibnamefont {Pani}}, \ and\
  \bibinfo {author} {\bibfnamefont {G.}~\bibnamefont {Raposo}},\ }\href@noop {}
  {\  (\bibinfo {year} {2021}{\natexlab{b}})},\ \Eprint
  {http://arxiv.org/abs/2105.06410} {arXiv:2105.06410 [gr-qc]} \BibitemShut
  {NoStop}%
\bibitem [{\citenamefont {{Ghez}}\ \emph {et~al.}(2009)\citenamefont {{Ghez}},
  \citenamefont {{Morris}}, \citenamefont {{Lu}}, \citenamefont {{Weinberg}},
  \citenamefont {{Matthews}}, \citenamefont {{Alexander}}, \citenamefont
  {{Armitage}}, \citenamefont {{Becklin}}, \citenamefont {{Brown}},
  \citenamefont {{Campbell}}, \citenamefont {{Do}}, \citenamefont {{Eckart}},
  \citenamefont {{Genzel}}, \citenamefont {{Gould}}, \citenamefont {{Hansen}},
  \citenamefont {{Ho}}, \citenamefont {{Lo}}, \citenamefont {{Loeb}},
  \citenamefont {{Melia}}, \citenamefont {{Merritt}}, \citenamefont
  {{Milosavljevic}}, \citenamefont {{Perets}}, \citenamefont {{Rasio}},
  \citenamefont {{Reid}}, \citenamefont {{Salim}}, \citenamefont
  {{Sch{\"o}del}},\ and\ \citenamefont {{Yelda}}}]{Ghez2009astro}%
  \BibitemOpen
  \bibfield  {author} {\bibinfo {author} {\bibfnamefont {A.}~\bibnamefont
  {{Ghez}}}, \bibinfo {author} {\bibfnamefont {M.}~\bibnamefont {{Morris}}},
  \bibinfo {author} {\bibfnamefont {J.}~\bibnamefont {{Lu}}}, \bibinfo {author}
  {\bibfnamefont {N.}~\bibnamefont {{Weinberg}}}, \bibinfo {author}
  {\bibfnamefont {K.}~\bibnamefont {{Matthews}}}, \bibinfo {author}
  {\bibfnamefont {T.}~\bibnamefont {{Alexander}}}, \bibinfo {author}
  {\bibfnamefont {P.}~\bibnamefont {{Armitage}}}, \bibinfo {author}
  {\bibfnamefont {E.}~\bibnamefont {{Becklin}}}, \bibinfo {author}
  {\bibfnamefont {W.}~\bibnamefont {{Brown}}}, \bibinfo {author} {\bibfnamefont
  {R.}~\bibnamefont {{Campbell}}}, \bibinfo {author} {\bibfnamefont
  {T.}~\bibnamefont {{Do}}}, \bibinfo {author} {\bibfnamefont {A.}~\bibnamefont
  {{Eckart}}}, \bibinfo {author} {\bibfnamefont {R.}~\bibnamefont {{Genzel}}},
  \bibinfo {author} {\bibfnamefont {A.}~\bibnamefont {{Gould}}}, \bibinfo
  {author} {\bibfnamefont {B.}~\bibnamefont {{Hansen}}}, \bibinfo {author}
  {\bibfnamefont {L.}~\bibnamefont {{Ho}}}, \bibinfo {author} {\bibfnamefont
  {F.}~\bibnamefont {{Lo}}}, \bibinfo {author} {\bibfnamefont {A.}~\bibnamefont
  {{Loeb}}}, \bibinfo {author} {\bibfnamefont {F.}~\bibnamefont {{Melia}}},
  \bibinfo {author} {\bibfnamefont {D.}~\bibnamefont {{Merritt}}}, \bibinfo
  {author} {\bibfnamefont {M.}~\bibnamefont {{Milosavljevic}}}, \bibinfo
  {author} {\bibfnamefont {H.}~\bibnamefont {{Perets}}}, \bibinfo {author}
  {\bibfnamefont {F.}~\bibnamefont {{Rasio}}}, \bibinfo {author} {\bibfnamefont
  {M.}~\bibnamefont {{Reid}}}, \bibinfo {author} {\bibfnamefont
  {S.}~\bibnamefont {{Salim}}}, \bibinfo {author} {\bibfnamefont
  {R.}~\bibnamefont {{Sch{\"o}del}}}, \ and\ \bibinfo {author} {\bibfnamefont
  {S.}~\bibnamefont {{Yelda}}},\ }in\ \href@noop {} {\emph {\bibinfo
  {booktitle} {astro2010: The Astronomy and Astrophysics Decadal Survey}}},\
  Vol.\ \bibinfo {volume} {2010}\ (\bibinfo {year} {2009})\ p.~\bibinfo {pages}
  {89},\ \Eprint {http://arxiv.org/abs/0903.0383} {arXiv:0903.0383
  [astro-ph.GA]} \BibitemShut {NoStop}%
\bibitem [{\citenamefont {Kormendy}\ and\ \citenamefont
  {Ho}(2013)}]{Kormendy_2013}%
  \BibitemOpen
  \bibfield  {author} {\bibinfo {author} {\bibfnamefont {J.}~\bibnamefont
  {Kormendy}}\ and\ \bibinfo {author} {\bibfnamefont {L.~C.}\ \bibnamefont
  {Ho}},\ }\href {\doibase 10.1146/annurev-astro-082708-101811} {\bibfield
  {journal} {\bibinfo  {journal} {Annual Review of Astronomy and Astrophysics}\
  }\textbf {\bibinfo {volume} {51}},\ \bibinfo {pages} {511} (\bibinfo {year}
  {2013})},\ \Eprint
  {http://arxiv.org/abs/https://doi.org/10.1146/annurev-astro-082708-101811}
  {https://doi.org/10.1146/annurev-astro-082708-101811} \BibitemShut {NoStop}%
\bibitem [{\citenamefont {{McConnell}}\ and\ \citenamefont
  {{Ma}}(2013)}]{McConnell2013ApJ}%
  \BibitemOpen
  \bibfield  {author} {\bibinfo {author} {\bibfnamefont {N.~J.}\ \bibnamefont
  {{McConnell}}}\ and\ \bibinfo {author} {\bibfnamefont {C.-P.}\ \bibnamefont
  {{Ma}}},\ }\href {\doibase 10.1088/0004-637X/764/2/184} {\bibfield  {journal}
  {\bibinfo  {journal} {\apj}\ }\textbf {\bibinfo {volume} {764}},\ \bibinfo
  {eid} {184} (\bibinfo {year} {2013})},\ \Eprint
  {http://arxiv.org/abs/1211.2816} {arXiv:1211.2816 [astro-ph.CO]} \BibitemShut
  {NoStop}%
\bibitem [{\citenamefont {Antonini}\ and\ \citenamefont
  {Perets}(2012)}]{Antonini:2012ad}%
  \BibitemOpen
  \bibfield  {author} {\bibinfo {author} {\bibfnamefont {F.}~\bibnamefont
  {Antonini}}\ and\ \bibinfo {author} {\bibfnamefont {H.~B.}\ \bibnamefont
  {Perets}},\ }\href {\doibase 10.1088/0004-637X/757/1/27} {\bibfield
  {journal} {\bibinfo  {journal} {Astrophys. J.}\ }\textbf {\bibinfo {volume}
  {757}},\ \bibinfo {pages} {27} (\bibinfo {year} {2012})},\ \Eprint
  {http://arxiv.org/abs/1203.2938} {arXiv:1203.2938 [astro-ph.GA]} \BibitemShut
  {NoStop}%
%%CITATION = ARXIV:1203.2938;%%
\bibitem [{\citenamefont {McKernan}\ \emph {et~al.}(2014)\citenamefont
  {McKernan}, \citenamefont {Ford}, \citenamefont {Kocsis}, \citenamefont
  {Lyra},\ and\ \citenamefont {Winter}}]{McKernan2012}%
  \BibitemOpen
  \bibfield  {author} {\bibinfo {author} {\bibfnamefont {B.}~\bibnamefont
  {McKernan}}, \bibinfo {author} {\bibfnamefont {K.~E.~S.}\ \bibnamefont
  {Ford}}, \bibinfo {author} {\bibfnamefont {B.}~\bibnamefont {Kocsis}},
  \bibinfo {author} {\bibfnamefont {W.}~\bibnamefont {Lyra}}, \ and\ \bibinfo
  {author} {\bibfnamefont {L.~M.}\ \bibnamefont {Winter}},\ }\href {\doibase
  10.1093/mnras/stu553} {\bibfield  {journal} {\bibinfo  {journal} {Monthly
  Notices of the Royal Astronomical Society}\ }\textbf {\bibinfo {volume}
  {441}},\ \bibinfo {pages} {900} (\bibinfo {year} {2014})},\ \Eprint
  {http://arxiv.org/abs/https://academic.oup.com/mnras/article-pdf/441/1/900/3018884/stu553.pdf}
  {https://academic.oup.com/mnras/article-pdf/441/1/900/3018884/stu553.pdf}
  \BibitemShut {NoStop}%
\bibitem [{\citenamefont {Chen}\ \emph {et~al.}(2019)\citenamefont {Chen},
  \citenamefont {Li},\ and\ \citenamefont {Cao}}]{Chen:2017xbi}%
  \BibitemOpen
  \bibfield  {author} {\bibinfo {author} {\bibfnamefont {X.}~\bibnamefont
  {Chen}}, \bibinfo {author} {\bibfnamefont {S.}~\bibnamefont {Li}}, \ and\
  \bibinfo {author} {\bibfnamefont {Z.}~\bibnamefont {Cao}},\ }\href {\doibase
  10.1093/mnrasl/slz046} {\bibfield  {journal} {\bibinfo  {journal} {Mon. Not.
  Roy. Astron. Soc.}\ }\textbf {\bibinfo {volume} {485}},\ \bibinfo {pages}
  {L141} (\bibinfo {year} {2019})},\ \Eprint {http://arxiv.org/abs/1703.10543}
  {arXiv:1703.10543 [astro-ph.HE]} \BibitemShut {NoStop}%
%%CITATION = ARXIV:1703.10543;%%
\bibitem [{\citenamefont {Inayoshi}\ \emph {et~al.}(2017)\citenamefont
  {Inayoshi}, \citenamefont {Tamanini}, \citenamefont {Caprini},\ and\
  \citenamefont {Haiman}}]{Inayoshi:2017hgw}%
  \BibitemOpen
  \bibfield  {author} {\bibinfo {author} {\bibfnamefont {K.}~\bibnamefont
  {Inayoshi}}, \bibinfo {author} {\bibfnamefont {N.}~\bibnamefont {Tamanini}},
  \bibinfo {author} {\bibfnamefont {C.}~\bibnamefont {Caprini}}, \ and\
  \bibinfo {author} {\bibfnamefont {Z.}~\bibnamefont {Haiman}},\ }\href
  {\doibase 10.1103/PhysRevD.96.063014} {\bibfield  {journal} {\bibinfo
  {journal} {Phys. Rev.}\ }\textbf {\bibinfo {volume} {D96}},\ \bibinfo {pages}
  {063014} (\bibinfo {year} {2017})},\ \Eprint
  {http://arxiv.org/abs/1702.06529} {arXiv:1702.06529 [astro-ph.HE]}
  \BibitemShut {NoStop}%
%%CITATION = ARXIV:1702.06529;%%
\bibitem [{\citenamefont {Chen}\ and\ \citenamefont
  {Han}(2018)}]{Chen:2018axp}%
  \BibitemOpen
  \bibfield  {author} {\bibinfo {author} {\bibfnamefont {X.}~\bibnamefont
  {Chen}}\ and\ \bibinfo {author} {\bibfnamefont {W.-B.}\ \bibnamefont {Han}},\
  }\href {\doibase 10.1038/s42005-018-0053-0} {\bibfield  {journal} {\bibinfo
  {journal} {Communications Physics}\ }\textbf {\bibinfo {volume} {1}},\
  \bibinfo {pages} {53} (\bibinfo {year} {2018})},\ \Eprint
  {http://arxiv.org/abs/1801.05780} {arXiv:1801.05780 [astro-ph.HE]}
  \BibitemShut {NoStop}%
%%CITATION = ARXIV:1801.05780;%%
\bibitem [{\citenamefont {{Tagawa}}\ \emph {et~al.}(2020)\citenamefont
  {{Tagawa}}, \citenamefont {{Haiman}},\ and\ \citenamefont
  {{Kocsis}}}]{Tagawa2020}%
  \BibitemOpen
  \bibfield  {author} {\bibinfo {author} {\bibfnamefont {H.}~\bibnamefont
  {{Tagawa}}}, \bibinfo {author} {\bibfnamefont {Z.}~\bibnamefont {{Haiman}}},
  \ and\ \bibinfo {author} {\bibfnamefont {B.}~\bibnamefont {{Kocsis}}},\
  }\href {\doibase 10.3847/1538-4357/ab9b8c} {\bibfield  {journal} {\bibinfo
  {journal} {\apj}\ }\textbf {\bibinfo {volume} {898}},\ \bibinfo {eid} {25}
  (\bibinfo {year} {2020})},\ \Eprint {http://arxiv.org/abs/1912.08218}
  {arXiv:1912.08218 [astro-ph.GA]} \BibitemShut {NoStop}%
\bibitem [{\citenamefont {Peng}\ and\ \citenamefont
  {Chen}(2021)}]{Peng&xian2021}%
  \BibitemOpen
  \bibfield  {author} {\bibinfo {author} {\bibfnamefont {P.}~\bibnamefont
  {Peng}}\ and\ \bibinfo {author} {\bibfnamefont {X.}~\bibnamefont {Chen}},\
  }\href {\doibase 10.1093/mnras/stab1419} {\  (\bibinfo {year} {2021}),\
  10.1093/mnras/stab1419},\ \Eprint {http://arxiv.org/abs/2104.07685}
  {arXiv:2104.07685 [astro-ph.HE]} \BibitemShut {NoStop}%
\bibitem [{\citenamefont {Cisneros}\ \emph {et~al.}(2015)\citenamefont
  {Cisneros}, \citenamefont {Goedecke}, \citenamefont {Beetle},\ and\
  \citenamefont {Engelhardt}}]{Cisneros:2012sk}%
  \BibitemOpen
  \bibfield  {author} {\bibinfo {author} {\bibfnamefont {S.}~\bibnamefont
  {Cisneros}}, \bibinfo {author} {\bibfnamefont {G.}~\bibnamefont {Goedecke}},
  \bibinfo {author} {\bibfnamefont {C.}~\bibnamefont {Beetle}}, \ and\ \bibinfo
  {author} {\bibfnamefont {M.}~\bibnamefont {Engelhardt}},\ }\href {\doibase
  10.1093/mnras/stv172} {\bibfield  {journal} {\bibinfo  {journal} {Mon. Not.
  Roy. Astron. Soc.}\ }\textbf {\bibinfo {volume} {448}},\ \bibinfo {pages}
  {2733} (\bibinfo {year} {2015})},\ \Eprint {http://arxiv.org/abs/1203.2502}
  {arXiv:1203.2502 [gr-qc]} \BibitemShut {NoStop}%
\bibitem [{\citenamefont {Meiron}\ \emph {et~al.}(2017)\citenamefont {Meiron},
  \citenamefont {Kocsis},\ and\ \citenamefont {Loeb}}]{Meiron:2016ipr}%
  \BibitemOpen
  \bibfield  {author} {\bibinfo {author} {\bibfnamefont {Y.}~\bibnamefont
  {Meiron}}, \bibinfo {author} {\bibfnamefont {B.}~\bibnamefont {Kocsis}}, \
  and\ \bibinfo {author} {\bibfnamefont {A.}~\bibnamefont {Loeb}},\ }\href
  {\doibase 10.3847/1538-4357/834/2/200} {\bibfield  {journal} {\bibinfo
  {journal} {Astrophys. J.}\ }\textbf {\bibinfo {volume} {834}},\ \bibinfo
  {pages} {200} (\bibinfo {year} {2017})},\ \Eprint
  {http://arxiv.org/abs/1604.02148} {arXiv:1604.02148 [astro-ph.HE]}
  \BibitemShut {NoStop}%
\bibitem [{\citenamefont {Randall}\ and\ \citenamefont
  {Xianyu}(2019{\natexlab{a}})}]{Randall:2018lnh}%
  \BibitemOpen
  \bibfield  {author} {\bibinfo {author} {\bibfnamefont {L.}~\bibnamefont
  {Randall}}\ and\ \bibinfo {author} {\bibfnamefont {Z.-Z.}\ \bibnamefont
  {Xianyu}},\ }\href {\doibase 10.3847/1538-4357/ab20c6} {\bibfield  {journal}
  {\bibinfo  {journal} {Astrophys. J.}\ }\textbf {\bibinfo {volume} {878}},\
  \bibinfo {pages} {75} (\bibinfo {year} {2019}{\natexlab{a}})},\ \Eprint
  {http://arxiv.org/abs/1805.05335} {arXiv:1805.05335 [gr-qc]} \BibitemShut
  {NoStop}%
\bibitem [{\citenamefont {Wong}\ \emph {et~al.}(2019)\citenamefont {Wong},
  \citenamefont {Baibhav},\ and\ \citenamefont {Berti}}]{Wong:2019hsq}%
  \BibitemOpen
  \bibfield  {author} {\bibinfo {author} {\bibfnamefont {K.~W.~K.}\
  \bibnamefont {Wong}}, \bibinfo {author} {\bibfnamefont {V.}~\bibnamefont
  {Baibhav}}, \ and\ \bibinfo {author} {\bibfnamefont {E.}~\bibnamefont
  {Berti}},\ }\href {\doibase 10.1093/mnras/stz2077} {\bibfield  {journal}
  {\bibinfo  {journal} {Mon. Not. Roy. Astron. Soc.}\ }\textbf {\bibinfo
  {volume} {488}},\ \bibinfo {pages} {5665} (\bibinfo {year} {2019})},\ \Eprint
  {http://arxiv.org/abs/1902.01402} {arXiv:1902.01402 [astro-ph.HE]}
  \BibitemShut {NoStop}%
\bibitem [{\citenamefont {Torres-Orjuela}\ \emph {et~al.}(2019)\citenamefont
  {Torres-Orjuela}, \citenamefont {Chen}, \citenamefont {Cao}, \citenamefont
  {Amaro-Seoane},\ and\ \citenamefont {Peng}}]{Torres-Orjuela:2018ejx}%
  \BibitemOpen
  \bibfield  {author} {\bibinfo {author} {\bibfnamefont {A.}~\bibnamefont
  {Torres-Orjuela}}, \bibinfo {author} {\bibfnamefont {X.}~\bibnamefont
  {Chen}}, \bibinfo {author} {\bibfnamefont {Z.}~\bibnamefont {Cao}}, \bibinfo
  {author} {\bibfnamefont {P.}~\bibnamefont {Amaro-Seoane}}, \ and\ \bibinfo
  {author} {\bibfnamefont {P.}~\bibnamefont {Peng}},\ }\href {\doibase
  10.1103/PhysRevD.100.063012} {\bibfield  {journal} {\bibinfo  {journal}
  {Phys. Rev. D}\ }\textbf {\bibinfo {volume} {100}},\ \bibinfo {pages}
  {063012} (\bibinfo {year} {2019})},\ \Eprint
  {http://arxiv.org/abs/1806.09857} {arXiv:1806.09857 [astro-ph.HE]}
  \BibitemShut {NoStop}%
\bibitem [{\citenamefont {Torres-Orjuela}\ \emph {et~al.}(2020)\citenamefont
  {Torres-Orjuela}, \citenamefont {Chen},\ and\ \citenamefont
  {Amaro-Seoane}}]{Torres-Orjuela:2020cly}%
  \BibitemOpen
  \bibfield  {author} {\bibinfo {author} {\bibfnamefont {A.}~\bibnamefont
  {Torres-Orjuela}}, \bibinfo {author} {\bibfnamefont {X.}~\bibnamefont
  {Chen}}, \ and\ \bibinfo {author} {\bibfnamefont {P.}~\bibnamefont
  {Amaro-Seoane}},\ }\href {\doibase 10.1103/PhysRevD.101.083028} {\bibfield
  {journal} {\bibinfo  {journal} {Phys. Rev. D}\ }\textbf {\bibinfo {volume}
  {101}},\ \bibinfo {pages} {083028} (\bibinfo {year} {2020})},\ \Eprint
  {http://arxiv.org/abs/2001.00721} {arXiv:2001.00721 [astro-ph.HE]}
  \BibitemShut {NoStop}%
\bibitem [{\citenamefont {D'Orazio}\ and\ \citenamefont
  {Loeb}(2020)}]{DOrazio:2019fbq}%
  \BibitemOpen
  \bibfield  {author} {\bibinfo {author} {\bibfnamefont {D.~J.}\ \bibnamefont
  {D'Orazio}}\ and\ \bibinfo {author} {\bibfnamefont {A.}~\bibnamefont
  {Loeb}},\ }\href {\doibase 10.1103/PhysRevD.101.083031} {\bibfield  {journal}
  {\bibinfo  {journal} {Phys. Rev. D}\ }\textbf {\bibinfo {volume} {101}},\
  \bibinfo {pages} {083031} (\bibinfo {year} {2020})},\ \Eprint
  {http://arxiv.org/abs/1910.02966} {arXiv:1910.02966 [astro-ph.HE]}
  \BibitemShut {NoStop}%
\bibitem [{\citenamefont {Ezquiaga}\ and\ \citenamefont
  {Zumalac\'arregui}(2020)}]{Ezquiaga:2020dao}%
  \BibitemOpen
  \bibfield  {author} {\bibinfo {author} {\bibfnamefont {J.~M.}\ \bibnamefont
  {Ezquiaga}}\ and\ \bibinfo {author} {\bibfnamefont {M.}~\bibnamefont
  {Zumalac\'arregui}},\ }\href {\doibase 10.1103/PhysRevD.102.124048}
  {\bibfield  {journal} {\bibinfo  {journal} {Phys. Rev. D}\ }\textbf {\bibinfo
  {volume} {102}},\ \bibinfo {pages} {124048} (\bibinfo {year} {2020})},\
  \Eprint {http://arxiv.org/abs/2009.12187} {arXiv:2009.12187 [gr-qc]}
  \BibitemShut {NoStop}%
\bibitem [{\citenamefont {Ezquiaga}\ \emph {et~al.}(2021)\citenamefont
  {Ezquiaga}, \citenamefont {Holz}, \citenamefont {Hu}, \citenamefont {Lagos},\
  and\ \citenamefont {Wald}}]{Ezquiaga:2020gdt}%
  \BibitemOpen
  \bibfield  {author} {\bibinfo {author} {\bibfnamefont {J.~M.}\ \bibnamefont
  {Ezquiaga}}, \bibinfo {author} {\bibfnamefont {D.~E.}\ \bibnamefont {Holz}},
  \bibinfo {author} {\bibfnamefont {W.}~\bibnamefont {Hu}}, \bibinfo {author}
  {\bibfnamefont {M.}~\bibnamefont {Lagos}}, \ and\ \bibinfo {author}
  {\bibfnamefont {R.~M.}\ \bibnamefont {Wald}},\ }\href {\doibase
  10.1103/PhysRevD.103.064047} {\bibfield  {journal} {\bibinfo  {journal}
  {Phys. Rev. D}\ }\textbf {\bibinfo {volume} {103}},\ \bibinfo {pages}
  {064047} (\bibinfo {year} {2021})},\ \Eprint
  {http://arxiv.org/abs/2008.12814} {arXiv:2008.12814 [gr-qc]} \BibitemShut
  {NoStop}%
\bibitem [{\citenamefont {Hoang}\ \emph {et~al.}(2018)\citenamefont {Hoang},
  \citenamefont {Naoz}, \citenamefont {Kocsis}, \citenamefont {Rasio},\ and\
  \citenamefont {Dosopoulou}}]{Hoang:2017fvh}%
  \BibitemOpen
  \bibfield  {author} {\bibinfo {author} {\bibfnamefont {B.-M.}\ \bibnamefont
  {Hoang}}, \bibinfo {author} {\bibfnamefont {S.}~\bibnamefont {Naoz}},
  \bibinfo {author} {\bibfnamefont {B.}~\bibnamefont {Kocsis}}, \bibinfo
  {author} {\bibfnamefont {F.~A.}\ \bibnamefont {Rasio}}, \ and\ \bibinfo
  {author} {\bibfnamefont {F.}~\bibnamefont {Dosopoulou}},\ }\href {\doibase
  10.3847/1538-4357/aaafce} {\bibfield  {journal} {\bibinfo  {journal} {The
  Astrophysical Journal}\ }\textbf {\bibinfo {volume} {856}},\ \bibinfo {pages}
  {140} (\bibinfo {year} {2018})},\ \Eprint {http://arxiv.org/abs/1706.09896}
  {arXiv:1706.09896 [astro-ph.HE]} \BibitemShut {NoStop}%
%%CITATION = ARXIV:1706.09896;%%
\bibitem [{\citenamefont {Randall}\ and\ \citenamefont
  {Xianyu}(2019{\natexlab{b}})}]{Randall:2019sab}%
  \BibitemOpen
  \bibfield  {author} {\bibinfo {author} {\bibfnamefont {L.}~\bibnamefont
  {Randall}}\ and\ \bibinfo {author} {\bibfnamefont {Z.-Z.}\ \bibnamefont
  {Xianyu}},\ }\href@noop {} {\bibfield  {journal} {\bibinfo  {journal}
  {arXiv:1902.08604}\ } (\bibinfo {year} {2019}{\natexlab{b}})},\ \Eprint
  {http://arxiv.org/abs/1902.08604} {arXiv:1902.08604 [astro-ph.HE]}
  \BibitemShut {NoStop}%
%%CITATION = ARXIV:1902.08604;%%
\bibitem [{\citenamefont {Fang}\ \emph {et~al.}(2019)\citenamefont {Fang},
  \citenamefont {Chen},\ and\ \citenamefont {Huang}}]{Fang_2019apj}%
  \BibitemOpen
  \bibfield  {author} {\bibinfo {author} {\bibfnamefont {Y.}~\bibnamefont
  {Fang}}, \bibinfo {author} {\bibfnamefont {X.}~\bibnamefont {Chen}}, \ and\
  \bibinfo {author} {\bibfnamefont {Q.-G.}\ \bibnamefont {Huang}},\ }\href
  {\doibase 10.3847/1538-4357/ab510e} {\bibfield  {journal} {\bibinfo
  {journal} {The Astrophysical Journal}\ }\textbf {\bibinfo {volume} {887}},\
  \bibinfo {pages} {210} (\bibinfo {year} {2019})}\BibitemShut {NoStop}%
\bibitem [{\citenamefont {Fang}\ and\ \citenamefont
  {Huang}(2020)}]{Fang_triple2020}%
  \BibitemOpen
  \bibfield  {author} {\bibinfo {author} {\bibfnamefont {Y.}~\bibnamefont
  {Fang}}\ and\ \bibinfo {author} {\bibfnamefont {Q.-G.}\ \bibnamefont
  {Huang}},\ }\href {\doibase 10.1103/PhysRevD.102.104002} {\bibfield
  {journal} {\bibinfo  {journal} {Phys. Rev. D}\ }\textbf {\bibinfo {volume}
  {102}},\ \bibinfo {pages} {104002} (\bibinfo {year} {2020})},\ \Eprint
  {http://arxiv.org/abs/2004.09390} {arXiv:2004.09390 [gr-qc]} \BibitemShut
  {NoStop}%
\bibitem [{\citenamefont {Yu}\ and\ \citenamefont {Chen}(2021)}]{Yuhang2020}%
  \BibitemOpen
  \bibfield  {author} {\bibinfo {author} {\bibfnamefont {H.}~\bibnamefont
  {Yu}}\ and\ \bibinfo {author} {\bibfnamefont {Y.}~\bibnamefont {Chen}},\
  }\href {\doibase 10.1103/PhysRevLett.126.021101} {\bibfield  {journal}
  {\bibinfo  {journal} {Phys. Rev. Lett.}\ }\textbf {\bibinfo {volume} {126}},\
  \bibinfo {pages} {021101} (\bibinfo {year} {2021})},\ \Eprint
  {http://arxiv.org/abs/2009.02579} {arXiv:2009.02579 [gr-qc]} \BibitemShut
  {NoStop}%
\bibitem [{\citenamefont {Cardoso}\ \emph
  {et~al.}(2021{\natexlab{a}})\citenamefont {Cardoso}, \citenamefont {Duque},\
  and\ \citenamefont {Khanna}}]{Cardoso_tuningfork2021}%
  \BibitemOpen
  \bibfield  {author} {\bibinfo {author} {\bibfnamefont {V.}~\bibnamefont
  {Cardoso}}, \bibinfo {author} {\bibfnamefont {F.}~\bibnamefont {Duque}}, \
  and\ \bibinfo {author} {\bibfnamefont {G.}~\bibnamefont {Khanna}},\ }\href
  {\doibase 10.1103/PhysRevD.103.L081501} {\bibfield  {journal} {\bibinfo
  {journal} {Phys. Rev. D}\ }\textbf {\bibinfo {volume} {103}},\ \bibinfo
  {pages} {L081501} (\bibinfo {year} {2021}{\natexlab{a}})},\ \Eprint
  {http://arxiv.org/abs/2101.01186} {arXiv:2101.01186 [gr-qc]} \BibitemShut
  {NoStop}%
\bibitem [{\citenamefont {{Teukolsky}}(1973)}]{Teukolsky1973}%
  \BibitemOpen
  \bibfield  {author} {\bibinfo {author} {\bibfnamefont {S.~A.}\ \bibnamefont
  {{Teukolsky}}},\ }\href {\doibase 10.1086/152444} {\bibfield  {journal}
  {\bibinfo  {journal} {\apj}\ }\textbf {\bibinfo {volume} {185}},\ \bibinfo
  {pages} {635} (\bibinfo {year} {1973})}\BibitemShut {NoStop}%
\bibitem [{\citenamefont {Chandrasekhar}(1993)}]{Chandrasekhar_1992book}%
  \BibitemOpen
  \bibfield  {author} {\bibinfo {author} {\bibfnamefont {S.}~\bibnamefont
  {Chandrasekhar}},\ }\href@noop {} {\emph {\bibinfo {title} {{The Mathematical
  Theory of Black Holes}}}}\ (\bibinfo  {publisher} {Clarendon Press, Oxford},\
  \bibinfo {year} {1993})\BibitemShut {NoStop}%
\bibitem [{\citenamefont {Mark}\ \emph {et~al.}(2017)\citenamefont {Mark},
  \citenamefont {Zimmerman}, \citenamefont {Du},\ and\ \citenamefont
  {Chen}}]{Mark:2017dnq}%
  \BibitemOpen
  \bibfield  {author} {\bibinfo {author} {\bibfnamefont {Z.}~\bibnamefont
  {Mark}}, \bibinfo {author} {\bibfnamefont {A.}~\bibnamefont {Zimmerman}},
  \bibinfo {author} {\bibfnamefont {S.~M.}\ \bibnamefont {Du}}, \ and\ \bibinfo
  {author} {\bibfnamefont {Y.}~\bibnamefont {Chen}},\ }\href {\doibase
  10.1103/PhysRevD.96.084002} {\bibfield  {journal} {\bibinfo  {journal} {Phys.
  Rev. D}\ }\textbf {\bibinfo {volume} {96}},\ \bibinfo {pages} {084002}
  (\bibinfo {year} {2017})},\ \Eprint {http://arxiv.org/abs/1706.06155}
  {arXiv:1706.06155 [gr-qc]} \BibitemShut {NoStop}%
\bibitem [{\citenamefont {M.L.Lidov}(1962)}]{Lidov1962}%
  \BibitemOpen
  \bibfield  {author} {\bibinfo {author} {\bibnamefont {M.L.Lidov}},\ }\href
  {\doibase 10.1016/0032-0633(62)90129-0} {\bibfield  {journal} {\bibinfo
  {journal} {Planetary and Space Science.}\ }\textbf {\bibinfo {volume} {9}},\
  \bibinfo {pages} {719} (\bibinfo {year} {1962})}\BibitemShut {NoStop}%
%%%
\bibitem [{\citenamefont {Kozai}(1962)}]{kozai1962}%
  \BibitemOpen
  \bibfield  {author} {\bibinfo {author} {\bibfnamefont {Y.}~\bibnamefont
  {Kozai}},\ }\href {\doibase 10.1086/108790} {\bibfield  {journal} {\bibinfo
  {journal} {Astron. J.}\ }\textbf {\bibinfo {volume} {67}},\ \bibinfo {pages}
  {591} (\bibinfo {year} {1962})}\BibitemShut {NoStop}%
%%CITATION = ANJOA,67,591;%%
\bibitem [{\citenamefont {Cardoso}\ \emph
  {et~al.}(2021{\natexlab{b}})\citenamefont {Cardoso}, \citenamefont {Duque},\
  and\ \citenamefont {Foschi}}]{Cardoso_lightring2021}%
  \BibitemOpen
  \bibfield  {author} {\bibinfo {author} {\bibfnamefont {V.}~\bibnamefont
  {Cardoso}}, \bibinfo {author} {\bibfnamefont {F.}~\bibnamefont {Duque}}, \
  and\ \bibinfo {author} {\bibfnamefont {A.}~\bibnamefont {Foschi}},\ }\href
  {\doibase 10.1103/PhysRevD.103.104044} {\bibfield  {journal} {\bibinfo
  {journal} {Phys. Rev. D}\ }\textbf {\bibinfo {volume} {103}},\ \bibinfo
  {pages} {104044} (\bibinfo {year} {2021}{\natexlab{b}})},\ \Eprint
  {http://arxiv.org/abs/2102.07784} {arXiv:2102.07784 [gr-qc]} \BibitemShut
  {NoStop}%
\bibitem [{\citenamefont {Finn}(1992)}]{Finn:1992wt}%
  \BibitemOpen
  \bibfield  {author} {\bibinfo {author} {\bibfnamefont {L.~S.}\ \bibnamefont
  {Finn}},\ }\href {\doibase 10.1103/PhysRevD.46.5236} {\bibfield  {journal}
  {\bibinfo  {journal} {Phys. Rev.}\ }\textbf {\bibinfo {volume} {D46}},\
  \bibinfo {pages} {5236} (\bibinfo {year} {1992})},\ \Eprint
  {http://arxiv.org/abs/gr-qc/9209010} {arXiv:gr-qc/9209010 [gr-qc]}
  \BibitemShut {NoStop}%
%%CITATION = GR-QC/9209010;%%
\bibitem [{\citenamefont {Lindblom}\ \emph {et~al.}(2008)\citenamefont
  {Lindblom}, \citenamefont {Owen},\ and\ \citenamefont
  {Brown}}]{Lindblom:2008cm}%
  \BibitemOpen
  \bibfield  {author} {\bibinfo {author} {\bibfnamefont {L.}~\bibnamefont
  {Lindblom}}, \bibinfo {author} {\bibfnamefont {B.~J.}\ \bibnamefont {Owen}},
  \ and\ \bibinfo {author} {\bibfnamefont {D.~A.}\ \bibnamefont {Brown}},\
  }\href {\doibase 10.1103/PhysRevD.78.124020} {\bibfield  {journal} {\bibinfo
  {journal} {Phys. Rev.}\ }\textbf {\bibinfo {volume} {D78}},\ \bibinfo {pages}
  {124020} (\bibinfo {year} {2008})},\ \Eprint {http://arxiv.org/abs/0809.3844}
  {arXiv:0809.3844 [gr-qc]} \BibitemShut {NoStop}%
%%CITATION = ARXIV:0809.3844;%%
\bibitem [{\citenamefont {Robson}\ \emph {et~al.}(2019)\citenamefont {Robson},
  \citenamefont {Cornish},\ and\ \citenamefont {Liug}}]{Cornish:2018dyw}%
  \BibitemOpen
  \bibfield  {author} {\bibinfo {author} {\bibfnamefont {T.}~\bibnamefont
  {Robson}}, \bibinfo {author} {\bibfnamefont {N.~J.}\ \bibnamefont {Cornish}},
  \ and\ \bibinfo {author} {\bibfnamefont {C.}~\bibnamefont {Liug}},\ }\href
  {\doibase 10.1088/1361-6382/ab1101} {\bibfield  {journal} {\bibinfo
  {journal} {Class. Quant. Grav.}\ }\textbf {\bibinfo {volume} {36}},\ \bibinfo
  {pages} {105011} (\bibinfo {year} {2019})},\ \Eprint
  {http://arxiv.org/abs/1803.01944} {arXiv:1803.01944 [astro-ph.HE]}
  \BibitemShut {NoStop}%
%%CITATION = ARXIV:1803.01944;%%
\bibitem [{\citenamefont {Apostolatos}\ \emph {et~al.}(1994)\citenamefont
  {Apostolatos}, \citenamefont {Cutler}, \citenamefont {Sussman},\ and\
  \citenamefont {Thorne}}]{Apostolatos:1994mx}%
  \BibitemOpen
  \bibfield  {author} {\bibinfo {author} {\bibfnamefont {T.~A.}\ \bibnamefont
  {Apostolatos}}, \bibinfo {author} {\bibfnamefont {C.}~\bibnamefont {Cutler}},
  \bibinfo {author} {\bibfnamefont {G.~J.}\ \bibnamefont {Sussman}}, \ and\
  \bibinfo {author} {\bibfnamefont {K.~S.}\ \bibnamefont {Thorne}},\ }\href
  {\doibase 10.1103/PhysRevD.49.6274} {\bibfield  {journal} {\bibinfo
  {journal} {Phys. Rev.}\ }\textbf {\bibinfo {volume} {D49}},\ \bibinfo {pages}
  {6274} (\bibinfo {year} {1994})}\BibitemShut {NoStop}%
%%CITATION = PHRVA,D49,6274;%%
\bibitem [{\citenamefont {Suzuki}\ \emph {et~al.}(2020)\citenamefont {Suzuki},
  \citenamefont {Nakamura},\ and\ \citenamefont {Yamada}}]{Suzuki:2020vfw}%
  \BibitemOpen
  \bibfield  {author} {\bibinfo {author} {\bibfnamefont {H.}~\bibnamefont
  {Suzuki}}, \bibinfo {author} {\bibfnamefont {Y.}~\bibnamefont {Nakamura}}, \
  and\ \bibinfo {author} {\bibfnamefont {S.}~\bibnamefont {Yamada}},\ }\href
  {\doibase 10.1103/PhysRevD.102.124063} {\bibfield  {journal} {\bibinfo
  {journal} {Phys. Rev. D}\ }\textbf {\bibinfo {volume} {102}},\ \bibinfo
  {pages} {124063} (\bibinfo {year} {2020})},\ \Eprint
  {http://arxiv.org/abs/2009.06999} {arXiv:2009.06999 [gr-qc]} \BibitemShut
  {NoStop}%
\bibitem [{BHP()}]{BHPtoolkit}%
  \BibitemOpen
  \href@noop {} {\enquote {\bibinfo {title} {{Black Hole Perturbation
  Toolkit}},}\ }\bibinfo {howpublished}
  {(\href{http://bhptoolkit.org/}{bhptoolkit.org})}\BibitemShut {NoStop}%
\bibitem [{cen()}]{centra}%
  \BibitemOpen
  \href@noop {} {\enquote {\bibinfo {title} {gravitation in tecnico},}\
  }\bibinfo {howpublished}
  {(\href{https://centra.tecnico.ulisboa.pt/network/grit/files/}{centra.tecnico.ulisboa.pt})}\BibitemShut
  {NoStop}%
\end{thebibliography}%

%%%%%%%%%%%%%%%%%%%%%%%%%%%%%%%%%%%%%%%%%%%%%%%%%%%%%%%%%%%%%%%%%%%%%%%%%%%%%%%%
\end{document}